\documentclass[
  aps,
  prd,
  a4paper,
  reprint,
  nofootinbib,
  superscriptaddress,
  floatfix,
  preprintnumbers
]{revtex4-2}

\usepackage[utf8]{inputenc}
\usepackage{microtype}

\usepackage{amsmath}
\usepackage{amsfonts}
\usepackage{amssymb}
\usepackage{bm}

\usepackage{graphicx}
\usepackage{longtable}
\usepackage{booktabs}
\usepackage{siunitx}
\usepackage[table]{xcolor}

\usepackage{soul}
\usepackage{xspace}
\usepackage[normalem]{ulem}
\usepackage{comment}
\usepackage{xurl}
\usepackage{lineno}

\usepackage{hyperref}

\hypersetup{
  colorlinks=true,
  linkcolor=blue,
  citecolor=blue,
  urlcolor=blue
}

\graphicspath{{figures/}}
\renewcommand{\arraystretch}{1.5}

\newcommand{\rd}{\textrm{d}}

\newcommand{\gagg}{\ensuremath{g_{a\gamma\gamma}}\xspace}

\begin{document}


\title{Searching for axions from the Galactic nuclear star cluster with NuSTAR}

\author{Ruben Zatini}\email{zatini@lapth.cnrs.fr}\email{rzatini@iac.es}
\affiliation{LAPTh, CNRS,  USMB, F-74940 Annecy, France}
\affiliation{Instituto de Astrof\'isica de Canarias,  C/ V\'ia L\'actea, s/n E38205 - La Laguna, Tenerife, Spain}
\affiliation{Universidad de La Laguna, Departamento de Astrof\'isica, La Laguna, Tenerife, Spain}
\author{Francesca Calore}\email{calore@lapth.cnrs.fr}
\affiliation{LAPTh, CNRS,  USMB, F-74940 Annecy, France}
\author{Jorge Martin Camalich}
\email{jcamalich@iac.es}
\affiliation{Instituto de Astrof\'isica de Canarias,  C/ V\'ia L\'actea, s/n E38205 - La Laguna, Tenerife, Spain}
\affiliation{Universidad de La Laguna, Departamento de Astrof\'isica, La Laguna, Tenerife, Spain}
\affiliation{CERN, Theoretical Physics Department, CH-1211 Geneva 23, Switzerland}
\author{Pasquale Dario Serpico}\email{serpico@lapth.cnrs.fr}
\affiliation{LAPTh, CNRS,  USMB, F-74940 Annecy, France}

\date{\today}

\begin{abstract}
The innermost region of the Milky Way hosts a young star cluster with a total mass exceeding $10^4\,  M_\odot$, including several tens of very massive stars ($M\gtrsim 60\,M_\odot$) whose core conditions favor the production of axions and axion-like particles via the Primakoff process. Emitted with energies of tens of keV, these exotic particles would efficiently convert into X-ray photons in the magnetised environments they cross, from the central parsecs around the Galactic Center to the local Galactic medium. We present the first calculation of this signal, relying on stellar simulations with the MESA stellar evolution code and state-of-the-art input concerning magnetic fields and gas densities along the line of sight. Further, we analyse NuSTAR data in the region of interest, focusing on the energy range 30--70 keV. 
We infer bounds on the coupling $\gagg\lesssim 3\times 10^{-12}\,$GeV$^{-1}$ for masses $m_a\lesssim 2\times 10^{-11}\,$eV, and limits   
$\gagg\lesssim 10^{-11}\,$GeV$^{-1}$ up to neV masses, comparable with the best bounds available in the literature, but with very different systematics (and systematics control).
\end{abstract}

\maketitle

\section{Introduction}
\label{sec:introduction}

Axions are hypothetical pseudoscalar particles that arise in the Peccei--Quinn solution to the strong CP problem in quantum chromodynamics (QCD)~\cite{Peccei:1977hh, Peccei:1977ur, Weinberg:1977ma, Wilczek:1977pj}. They are also well-motivated candidates for dark matter~\cite{Preskill:1982cy, Abbott:1982af, Dine:1982ah}. More generally, axion-like particles appear in many extensions of the Standard Model, including string theory~\cite{Svrcek:2006yi, Arvanitaki:2009fg}. 
The QCD axion naturally couples to photons, nucleons and, in some models, electrons, with strengths that depend on the axion decay constant $f_a$, which also sets the axion mass. For axion-like particles, instead, the mass and the couplings are independent parameters. We refer to both as axions throughout.

Stars provide natural laboratories for weakly coupled particles, and observations of their evolution and cooling have been widely used to constrain axion properties~\cite{Raffelt:1996wa, CaputoRaffelt2024}. In this work, we focus on searches for axions produced in stellar interiors through the Primakoff process, and subsequently converted into X-ray photons in astrophysical magnetic fields.
Both processes are governed by the axion--photon coupling $g_{a\gamma\gamma}$, which we constrain as a function of the axion mass, since also $m_a$ enters the conversion.
The rate of each process scales as $g_{a\gamma\gamma}^2$, the expected photon flux is proportional to $g_{a\gamma\gamma}^4$, and an upper limit on the flux constrains the coupling only through its fourth root. Improving the limit therefore requires much larger fluxes, i.e.~abundant axion sources in strongly magnetized environments.

Hot, massive stars can efficiently produce axions, with energies extending into the hard X-ray band. 
This has motivated searches with the NuSTAR X-ray telescope\footnote{\url{https://nustar.caltech.edu/}} towards the Quintuplet and Westerlund~1 super star clusters, which constrain $g_{a\gamma\gamma}\lesssim3.6\times10^{-12}\,\mathrm{GeV^{-1}}$ for $m_a\lesssim5\times10^{-11}\,$eV~\cite{DessertClusters2020}, and towards the galaxies M82 and M87, which give $g_{a\gamma\gamma}\lesssim6.4\times10^{-13}\,\mathrm{GeV^{-1}}$ for $m_a\lesssim10^{-10}\,$eV~\cite{NingSafdi2025}, both at $95\%$ confidence. These bounds complement those of the CAST helioscope~\cite{CAST2024} and of other astrophysical probes (see Ref.~\cite{AxionLimits} and Sec.~\ref{sec:results}). Searches based on conversion in large-scale magnetic fields, such as those of galaxies and galaxy clusters, lose sensitivity at higher masses: the conversion remains coherent only over the oscillation length $2 E/m_a^2$, which is already below a parsec for $m_a=10^{-9}\,$eV at tens of keV. Extending these searches to higher masses therefore requires a source embedded in strong magnetic fields on parsec scales.

The nuclear star cluster of the Milky Way offers a nearby target for such a search. Its young population in the central parsec, the young nuclear cluster (YNC), includes hot, massive stars and Wolf--Rayet stars~\cite{Paumard:2006, von_Fellenberg_2022}, while the surrounding Galactic Center (GC) environment contains strong magnetic fields~\cite{Ferriere2009}. The coexistence of an efficient axion source and a magnetized region on a parsec scale motivates a dedicated study of this target, which, to our knowledge, has not been considered before as a source of axions.

An additional novelty of this work is the modeling of the conversion along the whole path of the axions, from the structures of the central parsecs around Sgr~A$^\star$, through the Central Molecular Zone (CMZ) and the inner kiloparsec, to the large-scale field of the Galactic disk, including bound as well as free electrons in the effective photon mass. The predicted axion spectrum of the YNC peaks at about $50\,$keV, falling within the $3$--$79\,$keV band of NuSTAR. We therefore perform the first dedicated search for this signal in archival NuSTAR observations of the GC, and, in the absence of a  detection, set competitive bounds on the axion-photon coupling up to $10^{-8}$ eV in mass.

The paper is organized as follows. Section~\ref{sec:stellar_axion_emission} introduces the Primakoff emission and the stellar population model of the YNC. Section~\ref{sec:axion_photon_conversion} recalls the axion--photon conversion probability and the quantities that control it, and Sec.~\ref{sec:galactic_environment} describes the Galactic gas and magnetic field models and the resulting probability for the YNC. Section~\ref{sec:xray_analysis} derives the expected X-ray flux and describes the NuSTAR data and the statistical analysis. In the absence of an axion signal, we report, as main results, 95\% confidence level (C.L.) upper limits on the axion-photon coupling in Sec.~\ref{sec:results}. Results and implications are further discussed in Sec.~\ref{sec:conclusions}. Technical details of the stellar models, Galactic environment, data analysis and further sensitivity studies are collected in the appendices.

\section{Axion emission from the young nuclear cluster}
\label{sec:stellar_axion_emission}

The YNC hosts nearly 200 spectroscopically identified young stars, including OB supergiants, Wolf--Rayet stars, and OB main-sequence stars~\cite{von_Fellenberg_2022}, which makes it a powerful axion source.
Rather than by spectral type, we describe these stars by their initial mass and age, which determine their internal structure and hence their axion emission.
In this section, we describe the Primakoff production of axions in stellar interiors and the population model we adopt for the YNC, together with the resulting axion spectrum.

\subsection{Primakoff axion production in stars}
\label{sec:primakoff}

The axion--photon interaction is described by the effective term
\begin{equation}
    \label{eq:Lag_axion-photon}
    \mathcal{L}_{a\gamma} \supset
    -\frac{\gagg}{4}\,aF_{\mu\nu}\widetilde F^{\mu\nu}
    =\gagg\,a\,\mathbf E\cdot\mathbf B.
\end{equation}
Here $a$ is the axion field and $F_{\mu\nu}$ is the electromagnetic field-strength tensor associated to the electric field $\mathbf E$ and magnetic field $\mathbf B$. In a stellar plasma, this interaction allows a thermal photon to convert into an axion in the screened electric field of an ion or electron, known as Primakoff process.

Following Refs.~\cite{Raffelt1986, Carenza2025}, we write the axion energy luminosity per unit volume and axion energy as (natural units are used throughout)
\begin{equation}
    \label{eq:Primakoff_emissivity}
    \begin{split}
    \frac{\rd Q_a^{\rm P}}{\rd E}
    ={}&\frac{\gagg^{2}\,\kappa_s^{2}\,T\,E}{32\pi^{3}\left(e^{E/T}-1\right)}
    \sqrt{1-\frac{m_{\gamma}^{2}}{E^{2}}}\Theta(E-m_\gamma)\\
    &\times\left[
        \left(E^{2}+\frac{\kappa_s^{2}}{4}\right)
        \ln\!\left(1+\frac{4E^{2}}{\kappa_s^{2}}\right)-E^{2}
    \right]\,.
    \end{split}
\end{equation}
In this expression $T$ is the local temperature and $\kappa_s$ is the Debye screening scale, defined as
\begin{equation}
    \kappa_s^{2}=\frac{4\pi\alpha_{\rm em}}{T}
    \sum_i Z_i^{2}n_i\,.
\end{equation}
The sum runs over charged plasma constituents, both nuclear species of atomic number $Z_i$ and electrons, for which $Z_e=-1$, and $\alpha_{\rm em}$ is the fine-structure constant. Further,
  $m_\gamma$ is the effective photon mass, or photon plasma frequency $\omega_{\rm pl}$, given by
\begin{equation}
    \label{eq:eff_mass}
    m_\gamma= \sqrt{ \sum_i\dfrac{4\pi \alpha_{\rm em}Z_i^2 n_i}{m_i}}\simeq 3.69 \times 10^{-11}\,\mathrm{eV}\,\sqrt{\dfrac{n_e}{1\,\mathrm{cm^{-3}}}}\,.
\end{equation}
 Integrating the local emissivity over the stellar volume gives the differential axion luminosity,
\begin{equation}
    \label{eq:Luminosity_Primakoff}
    \frac{\rd L_a^{\star}}{\rd E}(E)
    =\int_{\rm star}\rd V\,\frac{\rd Q_a^{\rm P}}{\rd E}.
\end{equation}

Evaluating Eq.~\eqref{eq:Luminosity_Primakoff} requires the internal structure of the star, i.e. the profiles of temperature, density and chemical composition; the last two determine the number densities $n_i$ entering $\kappa_s$ and $m_\gamma$. These
profiles depend on the initial mass $M$ and on the age $t$ of the star. We
obtain them from stellar evolution simulations with the code
MESA~\cite{Paxton2011, Jermyn2023MESA}, as described in
Appendix~\ref{app:stellar_models}, for 32 initial masses between $20$ and $150\,M_\odot$, using the profiles saved along each evolutionary track.

\subsection{Population model and axion spectrum}
\label{sec:population_axion_spectrum}

Equation~\eqref{eq:Luminosity_Primakoff} gives the differential axion luminosity $\mathrm{d}L_a^\star/\mathrm{d}E(E;M,t)$ of a single star of initial mass $M$ and age $t$. We obtain the spectrum of the whole YNC by weighting these stellar spectra with the initial mass function (IMF) $\xi(M)$, and with a uniform distribution of stellar ages of width $\Delta t$ centered at $t_c$, which represents the finite duration of the star formation episode:
\begin{equation}
\label{eq:population_spectrum_main}
    \left\langle\frac{\rd L_a}{\rd E}\right\rangle(E)
    =\int\rd M\,\frac{\xi(M)}{\Delta t}
    \int_{t_c-\frac{\Delta t}{2}}^{t_c+\frac{\Delta t}{2}}\rd t\,
    \frac{\rd L_a^\star}{\rd E}(E;M,t),
\end{equation}
where $\xi(M)\propto M^{-\alpha}$ is normalized so that the stars with initial masses between 1 and $150\,M_\odot$ have a total mass $M_{\rm pop}$.
Details of the stellar models and population calculation are given in Appendix~\ref{app:stellar_models}.

For the fiducial population, we follow Ref.~\cite{Lu2013IMF}, which studied the young stars within a projected radius of $0.5\,$pc of Sgr~A$^\star$, which we adopt as the extent of the source region (Sec.~\ref{sec:source_average}). The authors found a top-heavy IMF, with $\alpha=1.7\pm0.2$, and a total initial stellar mass of $(1.4$--$3.7)\times10^4\,M_\odot$ in stars above $1\,M_\odot$; we adopt the midpoint of this range, $M_{\rm pop}=2.55\times10^4\,M_\odot$. Following their analysis, we assume a metallicity $Z=0.02$ and an age of $t_c=3.9\,$Myr, consistent with the best-fit age of $4\,$Myr of Ref.~\cite{Gallego-Cano}.

While the YNC is modeled as an instantaneous starburst
in Ref.~\cite{Lu2013IMF}, here we allow for a finite
formation episode and adopt $\Delta t=1\,$Myr; Ref.~\cite{Paumard:2006}
suggests a maximum duration of $2\,$Myr.

\begin{figure}
    \centering
    \includegraphics[width=\linewidth]{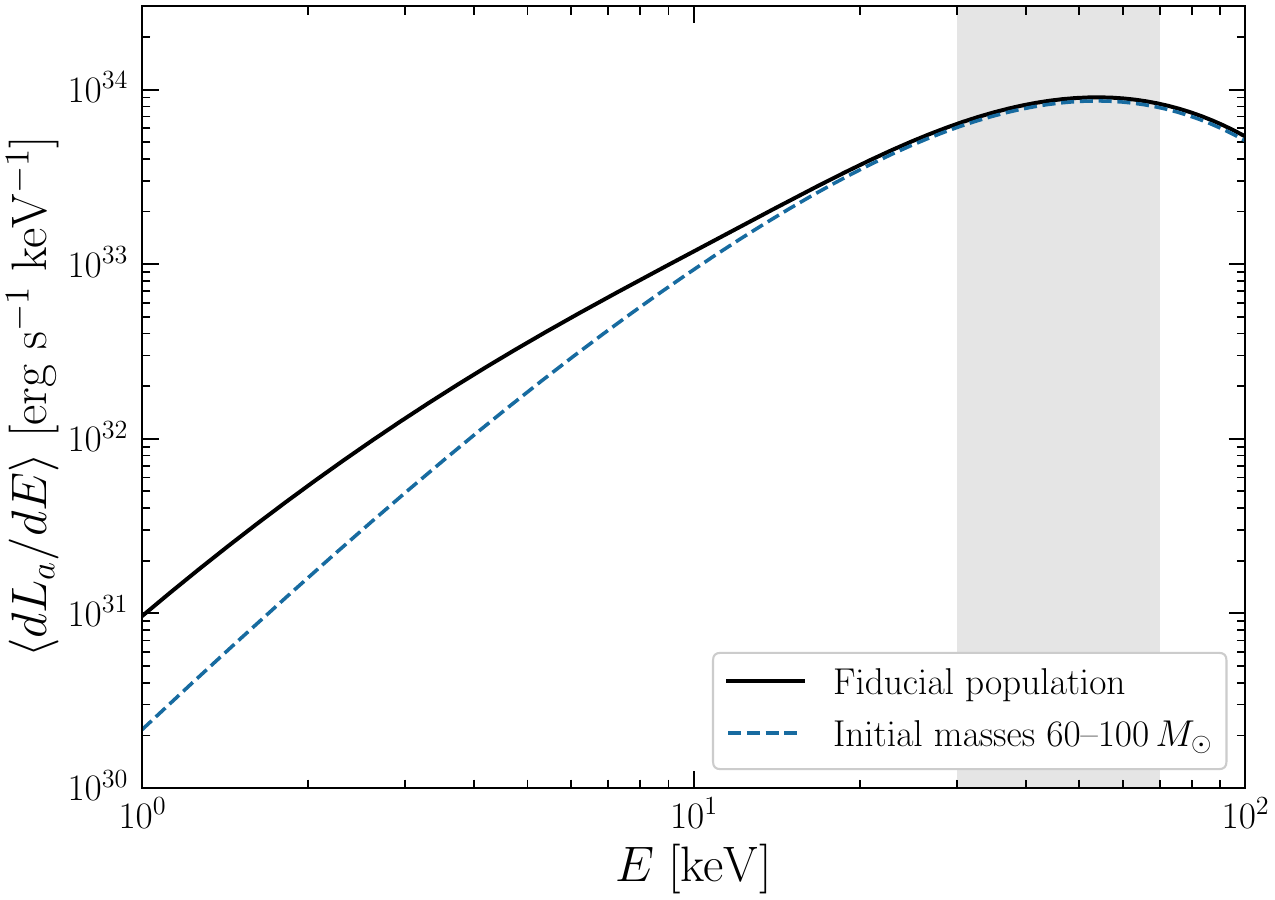}
    \caption{Differential axion luminosity of the YNC for
$g_{a\gamma\gamma}=10^{-12}\,\mathrm{GeV}^{-1}$ and the fiducial population
($\alpha=1.7$, $M_{\rm pop}=2.55\times10^4\,M_\odot$, $\Delta t=1\,$Myr, $t_c=3.9\,$Myr, $Z=0.02$; solid),
and contribution of the stars with initial masses between $60$ and
$100\,M_\odot$ (dashed). The shaded band indicates the $30$--$70\,$keV band of
our X-ray analysis. The spectra for other population parameters are shown in
Fig.~\ref{fig:population_variants}.}
    \label{fig:La_spectrum}
\end{figure}

Figure~\ref{fig:La_spectrum} shows the differential axion
luminosity of the YNC, which peaks at approximately
$50\,$keV with an amplitude of order
$10^{34}\,\mathrm{erg\,s^{-1}\,keV^{-1}}$ for
$g_{a\gamma\gamma}=10^{-12}\,\mathrm{GeV}^{-1}$.
We characterize the axion emission by its luminosity in the $30$--$70\,$keV band around the peak, which is also the band of our analysis (see Sec.~\ref{sec:xray_analysis}): $L_a\simeq3.4\times10^{35}\,(g_{a\gamma\gamma}/10^{-12}\,\mathrm{GeV}^{-1})^2\,$erg\,s$^{-1}$.

This emission is produced by evolved massive stars: during core hydrogen
burning, the $30$--$70\,$keV axion luminosity of a single star remains
below $\sim10^{30}\,$erg\,s$^{-1}$. After central hydrogen depletion, the
stellar interior heats up and the luminosity in this band rises by more
than four orders of magnitude (see Fig.~\ref{fig:mesa_single_star_stages}
for an $80\,M_\odot$ example). In the stars that dominate the emission,
these luminous phases last about $0.35\,$Myr, comparable to the burst
width, so that the adopted age distribution selects a narrow range of
initial masses: stars between $60$ and $100\,M_\odot$ provide about $95\%$
of $L_a$ (dashed curve in Fig.~\ref{fig:La_spectrum}). Less massive stars are still on the main sequence, while more
massive ones have already reached the end of their evolution
(Appendix~\ref{app:mesa_evolution}).

The spectrum depends on the population parameters mainly through its
normalization (Appendix~\ref{app:burst_average} and
Fig.~\ref{fig:population_variants}). The largest effect is that of
$M_{\rm pop}$, to which $L_a$ is proportional: the range of
Ref.~\cite{Lu2013IMF} changes it by about $\pm45\%$. Within the ranges
considered there, the IMF slope, the central age and the metallicity
change $L_a$ by at most $25\%$, $40\%$ and $25\%$, respectively.
Shorter bursts change it by only a few percent, so that our fiducial
estimate also represents an instantaneous burst, while a $2\,$Myr burst
reduces it by about $18\%$.

\section{Axion--photon conversion}
\label{sec:axion_photon_conversion}

In this section, we recall the formalism for the axion--photon conversion along a line of sight and the quantities that determine the conversion probability. The gas density and magnetic field of the Galaxy, and the resulting probability for the YNC, are described in Sec.~\ref{sec:galactic_environment}.

\subsection{Conversion along a line of sight}
\label{sec:conversion_los}

In the perturbative regime, where $P_{a\to\gamma}\ll1$, and neglecting birefringence and the attenuation of the photon component (the latter we apply after axion--photon conversion, see Sec.~\ref{sec:xray_flux}), the conversion probability can be written as~\cite{RaffeltStodolsky1988, MarshEtAl2022}
\begin{equation}
\label{eq:P_ag}
P_{a \to \gamma} \simeq \frac{g_{a\gamma\gamma}^2}{4}
\sum_{j=1,2}
\left|
\int_0^d \rd s\, B_j(s)\,
e^{i\Phi(s)}
\right|^2 ,
\end{equation}
where $B_j$ are the two magnetic-field components transverse to the line of sight, and
\begin{equation}
\label{eq:phase}
\Phi(s)\equiv\int_0^{s}\rd s'\,\left(\Delta_a-\Delta_\gamma(s')\right)
\end{equation}
is the phase accumulated from the source to $s$, with $\Delta_a=-m_a^2/(2E)$ and $\Delta_\gamma(s)=-m_\gamma^2(s)/(2E)$, $E$ the axion/photon energy and $m_\gamma$ the effective photon mass, given in Eq.~\eqref{eq:eff_mass}. Here $s=0$ identifies the source location, while  $s=d$ is at the observer (near Earth).

Evaluating Eq.~\eqref{eq:P_ag} requires the magnetic field and the interstellar gas along the path from the YNC to Earth: the transverse field drives the mixing, while the gas changes the accumulated phase $\Phi$ through the photon effective mass $m_\gamma$, which enters the term $\Delta_\gamma$.

\subsection{Effective photon mass}
\label{sec:gas_plasma}

The interstellar gas affects axion--photon conversion through its contribution to the photon dispersion relation. In particular, at energies well above the atomic binding energies, the effective mass of photons is given by Eq.~\eqref{eq:eff_mass}
where $n_e$ is the \emph{total} electron number density, including both free and bound electrons. The fact that, in the high-energy limit, bound electrons also contribute to the effective photon mass has been exploited in laboratory searches such as CAST to extend their sensitivity to higher axion masses \cite{E_Arik_2009}. In astrophysical searches for axion--photon conversion, including toward the GC region, the effective photon mass has instead been computed from the free-electron density of models of the ionized interstellar medium (e.g., Ref.~\cite{DessertClusters2020}). At X-ray energies, this underestimates $m_\gamma$ wherever the gas is mainly neutral. The difference is largest in the molecular gas of the CMZ and of the circumnuclear disk (CND), where free electrons are a small fraction of the total and where the magnetic field is strongest. The effective photon mass sets the phase in Eq.~\eqref{eq:P_ag}, and therefore affects both the coherence of the conversion and the masses at which resonances occur (Sec.~\ref{sec:conversion_regimes}). Including the bound electrons increases $m_\gamma$ and shifts these features to higher axion masses, which can change the limits derived from X-ray searches. We therefore estimate the total electron density from the interstellar hydrogen distribution, $n_{\rm H}=2n_{\rm H_2}+n_{\rm HI}+n_{\rm H^+}$, which is the quantity provided by the gas models of Sec.~\ref{sec:galactic_environment}. Each helium atom adds two electrons; for a helium abundance $n_{\rm He}/n_{\rm H}\simeq0.1$, close to the protosolar value~\cite{Asplund2009}, the total electron density is $n_e=n_{\rm H}+2n_{\rm He}\simeq1.2\,n_{\rm H}$. We therefore adopt $n_e=1.2\,n_{\rm H}$ throughout, neglecting the contribution of heavier elements.

\subsection{Coherent and resonant conversion}
\label{sec:conversion_regimes}

The behavior of Eq.~\eqref{eq:P_ag} depends on how fast $\Phi$ varies
along the path, and two limiting cases help to interpret our results.
Where $\Phi$ changes by much less than one radian over a length $L$, the
phase factor is nearly constant and the conversion is coherent; in the special case where the transverse field $B_\perp=(\sum_jB_j^2)^{1/2}$ is homogeneous over $L$, the probability reduces to the familiar expression $P_{a\to\gamma}\simeq(g_{a\gamma\gamma}B_\perp L/2)^2$. Wherever $m_\gamma\gg m_a$, $\Phi$ is set by the medium density alone, and the conversion there does not depend on $m_a$.

Where $\Phi$ varies rapidly, the integrand oscillates and averages out, suppressing the probability except near resonances at $s_k$, where the effective photon mass equals the
axion mass, $m_\gamma(s_k)=m_a$, and the phase is stationary,
${\rm d}\Phi/{\rm d}s=0$. If $B_\perp$ and ${\rm d}m_\gamma^2/{\rm d}s$ vary
little across a resonance and resonances do not overlap, the stationary-phase
approximation gives the probability of conversion at the resonance
$s_k$~\cite{CarenzaMarsh2023, UtrillaGines2024},
\begin{equation}
\label{eq:P_res}
P_k\simeq\frac{\pi g_{a\gamma\gamma}^2 B_\perp^2(s_k)\,E}
{\left|{\rm d}m_\gamma^2/{\rm d}s\right|_{s_k}},
\end{equation}
so that the conversion is strongest where the gas density varies slowly.

The two regimes also determine how the probability depends on energy, which
enters through the phase, $\Phi\propto E^{-1}$. In the coherent regime the phase
stays small and $P_{a\to\gamma}$ is nearly independent of $E$, while at a resonance
$P_k\propto E$. 

Between these limits, contributions from
different parts of the line of sight interfere with relative phases that scale as
$E^{-1}$, which produces oscillations of $P_{a\to\gamma}$ in both $E$ and $m_a$. When two resonances
separated by $\Delta s$ both contribute, the axion mass enters their relative phase
through $m_a^2\,\Delta s/(2E)$, and the probability oscillates in $m_a$ with a ``local period''
$\delta m_a=2\pi E/(m_a\,\Delta s)$. For resonances kiloparsecs apart this period
is short, and the positions of the oscillations depend on the gas column between
the resonances. The energy dependence
of the probability for the YNC therefore changes with the axion mass
(see Sec.~\ref{sec:Pbar}).

\subsection{Average over the source}
\label{sec:source_average}

Axions emitted at different points of an extended source reach the observer along different lines of sight, each with its own conversion probability. If all parts of the source emit with the same spectral shape, the expected photon flux is the spectrum of the whole source multiplied by the luminosity-weighted average of the conversion probability,
\begin{equation}
\label{eq:P_bar_general}
\bar P(E,m_a)=\frac{\int{\rm d}^3r\,\ell_a(\mathbf r)\,
P_{a\to\gamma}(E,m_a;\mathbf r)}{\int{\rm d}^3r\,\ell_a(\mathbf r)},
\end{equation}
where $\ell_a(\mathbf r)$ is the axion luminosity per unit volume at the position $\mathbf r$ in the source, and $P_{a\to\gamma}(E,m_a;\mathbf r)$ is given by Eq.~\eqref{eq:P_ag} along the line of sight from $\mathbf r$ to the observer. For a source much smaller than its distance, which is the case of the YNC, these lines of sight are parallel to a good approximation. The source model, $\ell_a(\mathbf r)$, for the YNC is specified in Sec.~\ref{sec:Pbar}.

\section{Galactic environment and conversion probability}
\label{sec:galactic_environment}

\begin{figure*}[t]
    \centering
    \includegraphics[width=\textwidth]{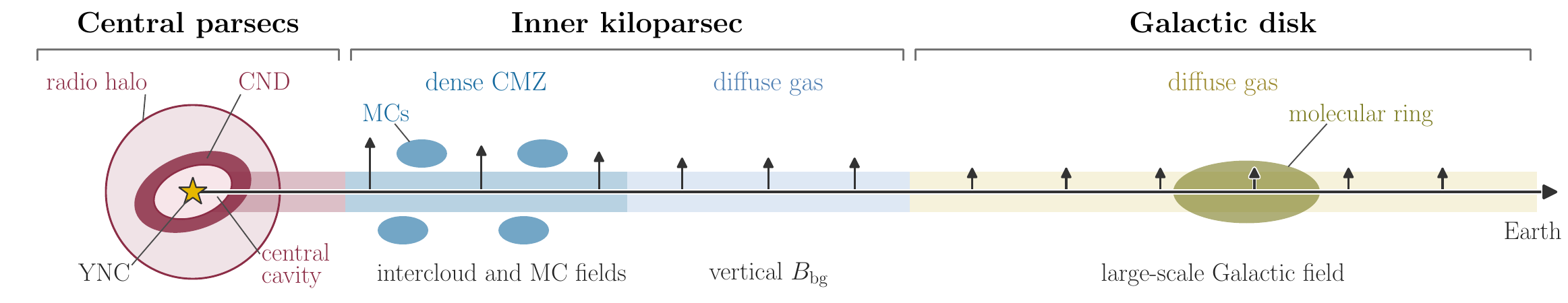}
    \caption{Sketch of the line of sight from the YNC to Earth, not to scale, with the three regions of Table~\ref{tab:gc_environment_models} and their main structures. The arrows represent the transverse magnetic field, and their length follows, qualitatively, its strength in Table~\ref{tab:gc_environment_models}.}
    \label{fig:los_schematic}
\end{figure*}

Galactic axion--photon conversion is commonly computed with large-scale models of the Galactic magnetic field, such as those of Jansson and Farrar (JF12)~\cite{JanssonFarrar2012} and Unger and Farrar (UF24)~\cite{UngerFarrar2024}. Neither describes the central region of the Galaxy: JF12, which we adopt on large scales, sets the field to zero within $1\,$kpc of the GC, and UF24 does not target this region either. Yet our axions are produced at the GC and cross the CMZ before reaching the large-scale field. There, the field is at least an order of magnitude stronger than in the Galactic disk and reaches mG strengths in molecular clouds (MCs), nonthermal filaments (NTFs) and the CND~\cite{Ferriere2009, Guenduez2016}. Using the model of Ref.~\cite{Guenduez2016}, Ref.~\cite{ZhuCMZ2025} found that this region dominates photon--axion conversion for the TeV source HESS~J1745$-$290. We therefore model the central parsecs and the inner kiloparsec explicitly. Table~\ref{tab:gc_environment_models} lists the three regions crossed by the line of sight, with their gas density, transverse field and the models we adopt, and Fig.~\ref{fig:los_schematic} shows their main structures.

\begin{table*}
    \caption{\label{tab:gc_environment_models}Regions of the line of sight of Fig.~\ref{fig:los_schematic}: range of the hydrogen density and of the transverse magnetic field along the central line of sight, and the gas and magnetic-field models we adopt. The modifications to the central field and the transitions between models are described in Sec.~\ref{sec:magnetic_field} and Appendix~\ref{app:galactic_center_models}.}
    \renewcommand{\arraystretch}{1.3}
    \setlength{\tabcolsep}{4pt}
    \small
    \newcommand{\gccell}[2]{\parbox[t]{#1\textwidth}{\raggedright #2}}
    \begin{tabular}{lccll}
        \toprule
        \gccell{0.16}{\textbf{Region}} & \boldmath$n_{\rm H}$ [cm$^{-3}$] & \boldmath$|B_\perp|$
        & \gccell{0.24}{\textbf{Gas}} & \gccell{0.28}{\textbf{Magnetic field}} \\
        \midrule
        \gccell{0.16}{\textbf{Central parsecs}\\ $r<10\,$pc} & $3$--$5\times10^5$ & up to $25\,$mG
        & \gccell{0.24}{Ferri\`ere (2012)~\cite{Ferriere2012}}
        & \gccell{0.28}{Guenduez et al.\ (2020)~\cite{Guenduez2016}, with confined CND field} \\
        \addlinespace
        \gccell{0.16}{\textbf{Inner kiloparsec}\\ $10\,\mathrm{pc}<r<1\,$kpc} & $0.05$--$2\times10^2$ & $10$--$60\,\mu$G
        & \gccell{0.24}{Ferri\`ere et al.\ (2007)~\cite{ferriere2007}, with the CMZ of Guenduez et al.\ (2020)~\cite{Guenduez2016}}
        & \gccell{0.28}{Guenduez et al.\ (2020)~\cite{Guenduez2016}: clouds, filaments, and intercloud field up to $r\simeq21\,$pc, then vertical $B_{\rm bg}$} \\
        \addlinespace
        \gccell{0.16}{\textbf{Galactic disk}\\ $r>1\,$kpc} & $0.05$--$7$ & $0.5$--$3\,\mu$G
        & \gccell{0.24}{Ferri\`ere et al.\ (2007)~\cite{ferriere2007} up to $3\,$kpc, Ferri\`ere (1998)~\cite{Ferriere1998} beyond}
        & \gccell{0.28}{Jansson \& Farrar (2012)~\cite{JanssonFarrar2012}} \\
        \bottomrule
    \end{tabular}
\end{table*}

\subsection{Gas and magnetic field along the line of sight}
\label{sec:magnetic_field}

We describe the three regions of Table~\ref{tab:gc_environment_models} from the source outward. For the gas we give the hydrogen density $n_{\rm H}$, with $n_e=1.2\,n_{\rm H}$ (Sec.~\ref{sec:gas_plasma}). Our implementation of the models, our modifications and the transitions between them are detailed in Appendix~\ref{app:galactic_center_models}.

\subsubsection{Central parsecs}

The YNC lies in the innermost $10\,$pc of the Galaxy, which we model as in Ref.~\cite{Ferriere2012}. Sgr~A$^\star$ sits inside a central cavity of ionized gas with $n_{\rm H}\sim10^3\,$cm$^{-3}$, surrounded by the CND, an inclined ring of molecular gas extending from about $1.2$ to $3\,$pc, with $n_{\rm H}\sim10^4$--$10^6\,$cm$^{-3}$. An extended radio halo and the Sgr~A~East supernova remnant (SNR) surround them with lower densities.

For the magnetic field of the central $200\,$pc, we adopt the model of Ref.~\cite{Guenduez2016}. At subparsec scales, its original CND field reaches hundreds of mG, whereas recent measurements indicate several mG in the CND~\cite{Guerra_2023} and 2--5 mG in the minispiral~\cite{Akshaya_2024}. We therefore confine the CND field to the inclined molecular ring, preserving $\nabla\cdot\mathbf B=0$, and let a fraction of it extend into the central cavity, where it gives fields of a few mG, as measured in the minispiral. Alternative choices of this confinement are discussed in Appendix~\ref{app:galactic_center_models}, and their effect on the limit is shown in Sec.~\ref{sec:results} and Appendix~\ref{app:sensitivity_tests}.

\subsubsection{Inner kiloparsec}

Within $\sim200\,$pc of the GC, the CMZ consists of dense molecular clouds embedded in a more diffuse medium. We adopt the gas model of Ref.~\cite{Guenduez2016}, based on Ref.~\cite{ferriere2007}, and its magnetic field, in which the MCs and the NTFs host fields up to the mG scale. Our lines of sight pass between the clouds (Fig.~\ref{fig:los_schematic}), through gas with $n_{\rm H}\lesssim2\times10^2\,$cm$^{-3}$ and $|B_\perp|\simeq10$--$60\,\mu$G. Beyond the CMZ and up to the molecular ring of the Galactic disk, they cross only diffuse, mostly ionized gas, with $n_{\rm H}\simeq0.05$--$0.1\,$cm$^{-3}$.

The field of the diffuse intercloud medium (ICM) of Ref.~\cite{Guenduez2016} decreases outward and falls to $10\,\mu$G at $r\simeq21\,$pc. From there to $1\,$kpc, where JF12 starts, we replace it with a homogeneous vertical background field, $B_{\rm bg}=10\,\mu$G. 
Estimates of the field on these scales range from $\sim10\,\mu$G, from the diffuse nonthermal radio emission~\cite{LaRosa_2005}, to the lower limit of $50\,\mu$G on $400\,$pc scales of Ref.~\cite{Crocker2010}. We adopt the reference field of Ref.~\cite{Guenduez2016}, $10\,\mu$G, as conservative benchmark, and consider $50\,\mu$G as the alternative case (Sec.~\ref{sec:Pbar}). 
We discuss the choice of $B_{\rm bg}$ in Sec.~\ref{sec:Pbar} and Appendix~\ref{app:galactic_center_models}.

\subsubsection{Galactic disk}

Between $1$ and $3\,$kpc the gas still follows Ref.~\cite{ferriere2007}, and beyond we adopt the axisymmetric model of Ref.~\cite{Ferriere1998}. The molecular gas forms a thin layer around the Galactic plane, peaking at $r\simeq4.5\,$kpc in the Galactic molecular ring, and the midplane density decreases from a few cm$^{-3}$ there to $n_{\rm H}\sim1\,$cm$^{-3}$ near the Sun. For $r>1\,$kpc we adopt the JF12 regular field, of a few $\mu$G, which comprises a spiral disk component, a toroidal halo and an X-shaped poloidal component.

Note that we conservatively neglect turbulent magnetic fields, which may increase the conversion at high masses~\cite{Fiorillo:2026rgo}.

\subsection{Conversion probability for the YNC}
\label{sec:Pbar}

Observations of the YNC constrain the distribution of its stars only in projection: both the IMF and the total initial stellar mass of Sec.~\ref{sec:population_axion_spectrum}~\cite{Lu2013IMF} and the surface-density profile $\Sigma$ of the young stars of the YNC in Ref.~\cite{Do2013} refer to the region within a projected radius of $0.5\,$pc of Sgr~A$^\star$, about $12.6''$ at the distance of the GC. We therefore describe the source as a disk of radius $R_{\rm p}=0.5\,$pc in the plane perpendicular to the GC--observer direction. 
Note that the whole young stellar population obeys the projected density profile $\Sigma\propto R^{-0.93\pm0.09}$~\cite{Do2013}, but the axion emission is dominated by a few tens of very massive evolved stars, which need not follow it. As a reference model, we therefore take $\ell_a$ uniform within the disk. Weighting the lines of sight with the observed profile instead changes the limit by at most $10\%$, depending on the mass (Appendix~\ref{app:sensitivity_tests}).
Since the position of the stars along the line of sight is poorly constrained, we place the disk in the plane of Sgr~A$^\star$, $x=0$, and show the effect on the limit of displacements of up to $\pm0.5\,$pc in  Appendix~\ref{app:sensitivity_tests}. As a result, Eq.~\eqref{eq:P_bar_general} becomes
\begin{equation}
\label{eq:P_bar}
    \bar P(E,m_a)=\frac{1}{\pi R_{\rm p}^2}
    \int_{R<R_{\rm p}}\!{\rm d}^2\mathbf R\,
    P_{a\to\gamma}(E,m_a;x_0,\mathbf R),
\end{equation}
where $\mathbf R=(y,z)$ is the projected position measured from
Sgr~A$^\star$, $R=|\mathbf R|$, and $x_0$ the position of the disk along the line of sight, so
that $\mathbf r=(x_0,\mathbf R)$ in Eq.~\eqref{eq:P_bar_general}. We take $x_0=0$
and compute $P_{a\to\gamma}$ with $d=D_{\rm GC}=8.178\,$kpc~\cite{GRAVITY2019}.

\begin{figure}
    \centering
    \includegraphics[width=\linewidth]{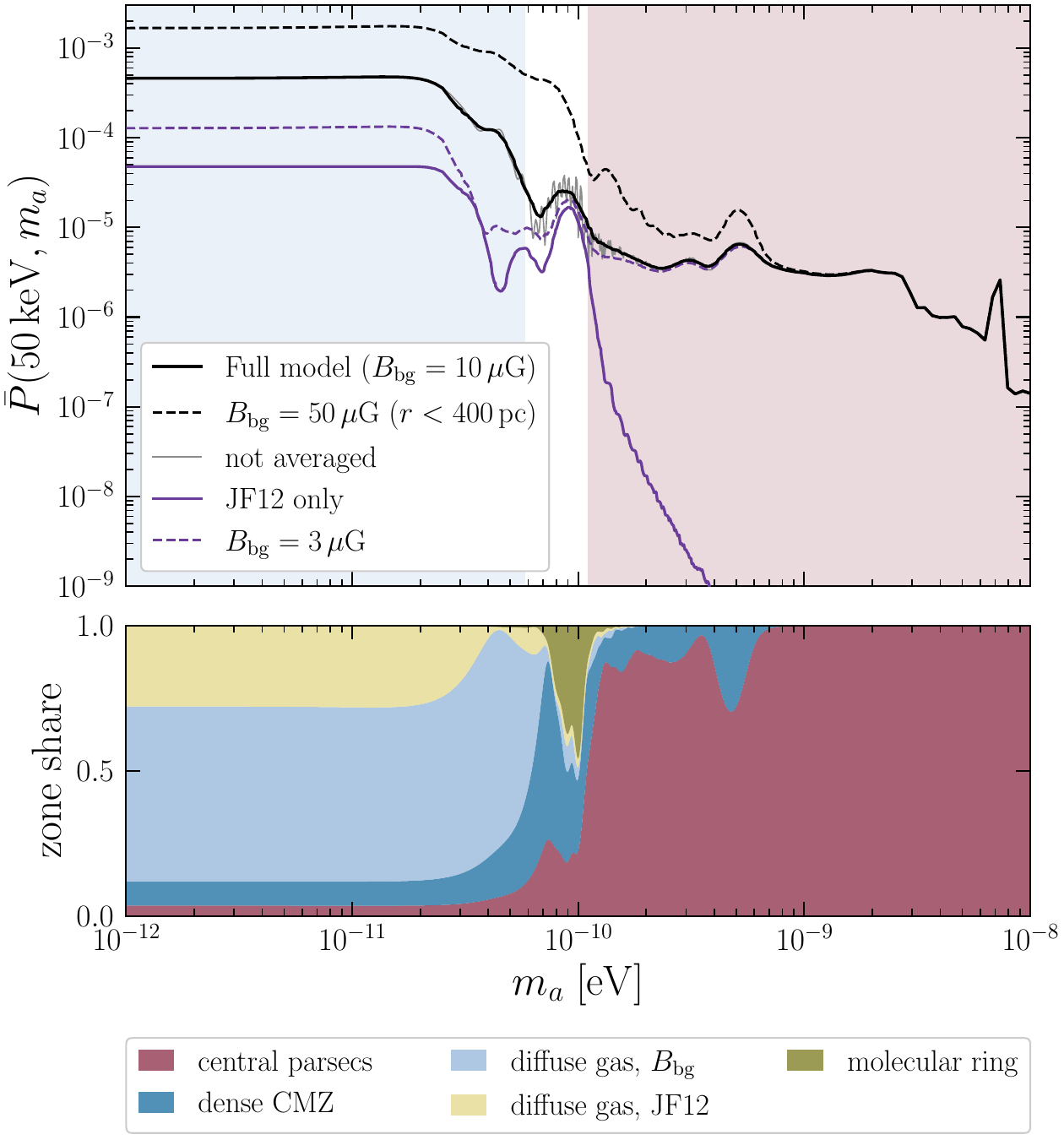}
    \caption{Disk-averaged axion--photon conversion probability at
$E=50\,\mathrm{keV}$ and $g_{a\gamma\gamma}=10^{-12}\,\mathrm{GeV^{-1}}$, for sources
uniformly distributed within a projected radius of $0.5\,$pc
in the plane $x=0$, and the zones of the line of sight that contribute to it.
Top: Mass dependence of the average probability under the full magnetic-field model (black) and JF12 field alone (purple), with the
same gas distribution. Black dashed: $50\,\mu$G within $r\simeq400\,$pc, the lower limit of Ref.~\cite{Crocker2010}. Purple dashed: $B_{\rm bg}=3\,\mu$G, the transverse JF12 field at $1\,$kpc extended into the inner kiloparsec. For $m_a\le10^{-9}\,$eV, all curves are averaged over $\pm0.05$ in $\log_{10}m_a$ (Sec.~\ref{sec:Pbar}). Thin gray: full model before averaging. Bottom: relative contribution of each zone of Fig.~\ref{fig:los_schematic} (boundaries at
$r=10\,$pc, $300\,$pc, $1\,$kpc and $3$--$6\,$kpc) to the conversion along the central
line of sight, from the integral of Eq.~\eqref{eq:P_ag} evaluated within each
zone, without the interference between zones. The shading in the top panel
marks the zone that contributes more than half (unshaded: none).}
    \label{fig:Pbar}
\end{figure}

We illustrate the mass dependence of the conversion probability at
$E=50\,$keV, near the peak of the axion luminosity spectrum of our adopted
stellar population (Sec.~\ref{sec:population_axion_spectrum}). The top panel of
Fig.~\ref{fig:Pbar} compares $\bar P(50\,\mathrm{keV},m_a)$ for our full
magnetic-field model and for the JF12 field alone, with the same gas
distribution. The bottom panel shows which zone of the line of sight provides the
conversion at each mass. From this figure, we can appreciate the relevance of 
including the central magnetic field in our model, which provides most of the conversion above $10^{-10}\,$eV in mass.

In particular, below $m_a\simeq2.5\times10^{-11}\,$eV, the conversion at $50\,$keV is coherent over kiloparsec scales, and both models show a plateau. The full model predicts $\bar P\simeq5\times10^{-4}\, (g_{a\gamma\gamma}/10^{-12}\,\mathrm{GeV^{-1}})^2$,
about ten times the JF12 value, because it adds the conversion in the diffuse gas of the inner kiloparsec, driven by $B_{\rm bg}=10\,\mu$G, our fiducial value (Sec.~\ref{sec:magnetic_field}).
The plateau therefore scales with the adopted value of $B_{\rm bg}$: adopting the lower limit of Ref.~\cite{Crocker2010}, $50\,\mu$G within $r\simeq400\,$pc and $10\,\mu$G up to $1\,$kpc (black dashed line in Fig.~\ref{fig:Pbar}), the plateau rises by a factor of $3.6$ and the low-mass limit on $g_{a\gamma\gamma}$ becomes stronger by a factor of $1.39$, whereas above $10^{-9}\,$eV, the limit is unchanged. Our fiducial value is thus conservative. For comparison, the purple dashed curve extends into the inner kiloparsec the transverse JF12 field at $1\,$kpc, $B_{\rm bg}=3\,\mu$G (Fig.~\ref{fig:field_join}), below all estimates of the field strength in this region. In that case, the plateau decreases by a factor of $3.6$ and the low-mass limit weakens by a factor of $1.38$, while above $3\times10^{-10}\,$eV the two limits differ by $\lesssim2\%$.

Between $6\times10^{-11}$ and $1.1\times10^{-10}\,$eV, resonances (Eq.~\eqref{eq:P_res}) at the edge of the dense CMZ and in the molecular ring contribute together with the central parsecs. Their interference produces the oscillations of the gray curve, as anticipated in Sec.~\ref{sec:conversion_regimes}. The positions of these oscillations are not a robust prediction: the relative phase is set by the gas column between the GC and the molecular ring, and the corresponding gas density uncertainties shift the phase by more than $2\pi$. For $m_a\le10^{-9}\,$eV we therefore average $\bar P$ uniformly in $\log_{10}m_a$ over $\pm0.05$ around each mass, as displayed by the black curve. For resonances near the GC and in the molecular ring, $\Delta s\simeq6\,$kpc, this window contains at least one oscillation period over $30$--$70\,$keV.

Above $1.1\times10^{-10}\,$eV, the JF12-only probability is strongly
suppressed: $m_\gamma$ in the Galactic disk reaches at most
$\simeq1.04\times10^{-10}\,$eV, in the molecular ring, so no resonance
remains along the large-scale field.
The full magnetic field model instead shows a second plateau,
$\bar P\simeq3$--$4\times10^{-6}$ up to $\simeq2.5\times10^{-9}\,$eV, which comes from coherent conversion in the mG field of the central cavity: this field is
confined to small scales, over which the accumulated phase remains small, so
that $\bar P$ depends little on $m_a$ and on $E$. On top of it, resonances in the
dense CMZ, whose density decreases slowly, produce the bump near
$5\times10^{-10}\,$eV. At higher masses, resonances in the cavity and in the CND
keep $\bar P$ of order $10^{-6}$ up to $\simeq7\times10^{-9}\,$eV.
For a source at the GC, the conversion is thus controlled by the inner Galaxy rather than by the large-scale field: the inner kiloparsec sets the low-mass plateau, and the CMZ and the central parsecs provide essentially all the conversion above $\simeq10^{-10}\,$eV. The prediction is therefore most sensitive to the fields in these regions: $B_{\rm bg}$ on the plateau and, at higher masses, the field in the central cavity, whose strength changes $\bar P$ by up to a factor of four between $10^{-10}$ and $2\times10^{-9}\,$eV (Appendix~\ref{app:galactic_center_models}).

\begin{figure}
    \centering
    \includegraphics[width=\linewidth]{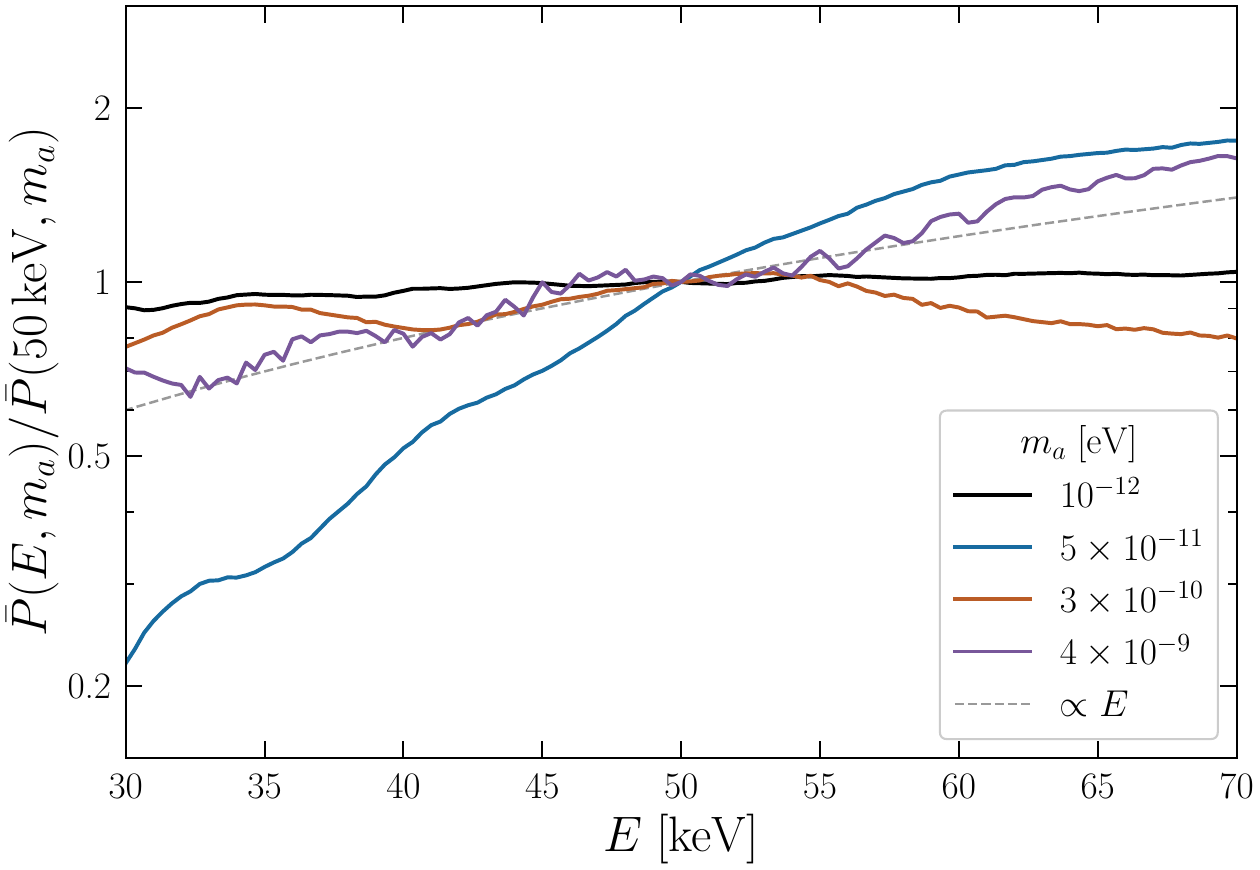}
    \caption{Energy dependence of the disk-averaged conversion probability in
    the $30$--$70\,$keV band, normalized to its value at $50\,$keV, for four axion
    masses representative of the regimes of Fig.~\ref{fig:Pbar}: the low-mass
    plateau ($10^{-12}\,$eV), the loss of coherence on kiloparsec scales
    ($5\times10^{-11}\,$eV), the conversion in the CMZ and in the central cavity
    ($3\times10^{-10}\,$eV), and the resonances in the central parsecs
    ($4\times10^{-9}\,$eV). For $m_a\le10^{-9}\,$eV, the curves are averaged over
    $\pm0.05$ in $\log_{10}m_a$, as in Fig.~\ref{fig:Pbar}. The dashed gray line is
    proportional to $E$.}
    \label{fig:Pbar_energy}
\end{figure}

The probability also depends on energy, as anticipated in Sec.~\ref{sec:conversion_regimes}. Figure~\ref{fig:Pbar_energy} shows this dependence across the $30$--$70\,$keV band for four representative masses. The average over the source and the logarithmic average in $m_a$ smooth the features of individual lines of sight, but a residual dependence remains. On the low-mass plateau, where the conversion is coherent, $\bar P$ varies by less than $20\%$ below $1.5\times10^{-11}\,$eV. Near $5\times10^{-11}\,$eV, where the coherence on kiloparsec scales is lost, it varies by up to a factor of $9$, and between
$1.2\times10^{-10}$ and $2\times10^{-9}\,$eV by less than a factor of $2$. Above $3\times10^{-9}\,$eV, where resonances in the central cavity and in the CND dominate, it grows with $E$, on average roughly linearly, as expected from Eq.~\eqref{eq:P_res}. We therefore adopt $\bar P(E,m_a)$ with its full energy dependence to compute the axion-induced X-ray signal (Sec.~\ref{sec:xray_flux}).

\section{X-ray flux and data analysis}
\label{sec:xray_analysis}

Combining the population spectrum $\langle\mathrm{d}L_a/\mathrm{d}E\rangle$ of
the YNC (Sec.~\ref{sec:population_axion_spectrum}) with the disk-averaged conversion
probability (Sec.~\ref{sec:Pbar}), we obtain the axion-induced photon flux at
Earth, integrated over the projected source and before interstellar
attenuation,
\begin{equation}
    \label{eq:Xray_flux}
    \begin{split}
    \phi_a(E;m_a)
    &=\frac{1}{4\pi D_{\rm GC}^2\,E}
    \left\langle \frac{\mathrm{d}L_a}{\mathrm{d}E}\right\rangle(E)\,
    \bar P(E,m_a)\\[2pt]
    &=\mu\,\phi_{a,0}(E;m_a),
    \end{split}
\end{equation}
where $\phi_{a,0}$ is evaluated at the reference coupling
$g_0=10^{-12}\,\mathrm{GeV}^{-1}$ and $\mu=(g_{a\gamma\gamma}/g_0)^4$ is the
signal strength, so that an upper limit on $\mu$ constrains $g_{a\gamma\gamma}$ only through its fourth root. The quartic dependence follows from the quadratic scaling of
both Primakoff production and perturbative conversion. For the fiducial population and magnetic field model (Secs.~\ref{sec:population_axion_spectrum} and \ref{sec:galactic_environment}), at $E=50\,$keV and $m_a=10^{-12}\,$eV, we find
\begin{equation}
    \label{eq:reference_flux}
    \phi_{a,0}\simeq6.4\times10^{-9}\,
    \mathrm{cm^{-2}\,s^{-1}\,keV^{-1}}.
\end{equation}

In the following, we describe the NuSTAR data (Sec.~\ref{sec:nustar_data}), the model of the spectrum measured in the source region (Sec.~\ref{sec:xray_flux}), and the statistical analysis procedure we followed to obtain the final limits (Sec.~\ref{sec:nustar_analysis}).

\subsection{NuSTAR observations and data reduction}
\label{sec:nustar_data}

We search for axion-induced X-ray emission in 11 archival NuSTAR observations of the GC taken between July 2012 and May 2021 (Appendix~\ref{app:nustar_data}). We give priority to long observations covering Sgr~A$^\star$, and add shorter ones with a favorable pointing geometry, requiring that the source and background regions defined below are fully covered, as checked on the exposure maps of each focal plane module. The two focal plane modules (FPMA and FPMB) of each observation are treated as separate datasets, for a total exposure of 1.65 Ms. We use the HEASoft/NuSTARDAS software package (HEASoft 6.37.1, NuSTARDAS 2.1.6, CALDB 20260903) to reduce the data, following the standard procedures of the NuSTAR pipeline.
For observations affected by flares from Sgr~A$^\star$, we exclude the corresponding time intervals documented in the literature (Appendix~\ref{app:nustar_data}). Source spectra are extracted from a disk of radius $30''$ centered on Sgr~A$^\star$ (ON
region), which contains the projected YNC disk of radius $12.6''$ and about half
of the NuSTAR point spread function (PSF). Background spectra are extracted from an annulus between $90''$ and $120''$ in the same observation (OFF region), chosen close to the ON region, to sample the same diffuse emission, but far enough to contain only about $5\%$ of the PSF of a source at Sgr~A$^\star$. The OFF spectra are rescaled to the ON region by the ratio of exposures and extraction areas. We treat the choice of the OFF region and the residual nonuniformity of the annulus as a systematic uncertainty, which we assess by using each of its four quadrants as OFF region. Their effect on the limit is discussed in Appendix~\ref{app:sensitivity_tests}. For each module and observation we also compute the ancillary response file (ARF) and the redistribution matrix file (RMF), which describe the effective area and the energy redistribution of each dataset (see Eq.~\eqref{eq:expected_counts}). Details of the observations, of the data reduction and of the background regions are given in Appendix~\ref{app:data_analysis}.

\subsection{Signal and continuum model}
\label{sec:xray_flux}

We restrict the analysis to measured energies between $30$ and $70\,$keV, in eight bins of $5\,$keV, where the NuSTAR effective area overlaps with the peak of the expected signal. The ON region also contains hard X-ray emission from other sources in the central parsec: above $20\,$keV
it is dominated by the pulsar-wind-nebula candidate G359.95$-$0.04 and by
unresolved diffuse emission concentrated in the inner few
parsecs~\cite{Mori2015,Zhang2017}. In the phenomenological model of
Ref.~\cite{Zhang2017} for the same region, the thermal components become
negligible above $20\,$keV, where the power-law component dominates. We
therefore model the emission in our band with a power law, while at lower energies a single power law may not describe correctly the emission of the source region, which includes components with different origin and spectra. On the other hand, above $70\,$keV the effective area falls steeply. We show the effect of lower edges of $25$ and $35\,$keV on the limit in Appendix~\ref{app:sensitivity_tests}.

We apply the same
interstellar transmission to the signal and to the continuum,\footnote{For the
signal this is conservative: photons converted along the line of sight cross
only part of the column.} so that the photon flux incident on the telescope is
\begin{equation}
\label{eq:incident_flux}
\phi(E)=\mathcal{T}(E)\left[\mu\,\phi_{a,0}(E;m_a)
+K\left(\frac{E}{40\,\mathrm{keV}}\right)^{-\Gamma}\right],
\end{equation}
where $K$ is the photon flux density of the continuum at $40\,$keV and $\mathcal{T}$ is
the interstellar transmission. The continuum thus describes the sources in the
ON region, while the instrumental background and the large-scale diffuse
emission are measured in a separate OFF region (Sec.~\ref{sec:nustar_data}).

In our band the interstellar attenuation is dominated by Compton
scattering out of the line of sight, while photoelectric absorption is
negligible. The transmission is therefore
\begin{equation}
\label{eq:transmission}
\mathcal{T}(E)=\exp\!\left(-1.2\,N_{\rm H}\,\sigma_{\rm KN}(E)\right),
\end{equation}
where $\sigma_{\rm KN}$ is the Klein--Nishina cross section per electron, the
factor $1.2$ counts the electrons per hydrogen atom, as in
Sec.~\ref{sec:gas_plasma}, and $N_{\rm H}$ is the hydrogen column density
between Earth and the YNC. We fix the hydrogen column to $N_{\rm H}=1.66\times10^{23}\,\mathrm{cm^{-2}}$, measured with NuSTAR toward Sgr~A$^\star$ from a joint fit of four flares~\cite{Barriere2014}, whose time intervals are excluded from our analysis. We checked that varying $N_{\rm H}$ within the $90\%$ C.L.~range of the same reference, $(1.05$--$2.36)\times10^{23}\,\mathrm{cm^{-2}}$, changes the limit by at most $1.2\%$. With the fiducial choice for the column, the Compton optical depth, $\tau(E)=1.2\,N_{\rm H}\,\sigma_{\rm KN}(E)$, decreases from $0.12$ at $30\,$keV to $0.11$ at $70\,$keV, and the transmission in Eq.~\eqref{eq:transmission} reduces the expected signal by about $11\%$.

The expected source counts in the
measured-energy bin $i$ of dataset $d$ are
\begin{equation}
\label{eq:expected_counts}
M_{di}=t_d\int\mathrm{d}E\,A_d(E)\,R_{di}(E)\,\phi_d(E),
\end{equation}
where $t_d$ is the exposure, $A_d$ the effective area from the ARF, which includes the fraction of the PSF enclosed in the ON region for a point source at Sgr~A$^\star$,\footnote{We approximate the YNC, a disk of radius $12.6''$, as a point source: this overestimates the fraction of its PSF in the ON region by about $3\%$. We estimate the impact on the final limit to be less than $1\%$.} $R_{di}$ the probability, from the RMF, that a photon of energy $E$ is detected in bin
$i$, and $\phi_d$ is Eq.~\eqref{eq:incident_flux} with a continuum normalization $K_d$ for each dataset. The signal strength $\mu$ and the photon index $\Gamma$ are common to all datasets, while the separate normalizations absorb the variability of the source region between observations.

\subsection{Spectral analysis framework and upper limits}
\label{sec:nustar_analysis}

For each axion mass, we compare the ON and OFF spectra of all datasets with the model of Eq.~\eqref{eq:incident_flux}, folded through the instrumental response, and derive a one-sided 95\% C.L.~upper limit on $g_{a\gamma\gamma}$ from the profile likelihood ratio.

We keep the 22 datasets separate, so that each spectrum retains its own response and continuum normalization, and therefore fit them with a joint Poisson likelihood. The ON counts $n_{di}$ and the OFF counts $m_{di}$ are independent Poisson variables. With $b_{di}$ the expected OFF counts and $\alpha_d$ the ON/OFF scaling factor, the likelihood is
\begin{equation}
\label{eq:likelihood}
\begin{split}
\mathcal L={}&\prod_{d,i}
\operatorname{Pois}\!\left(n_{di}\mid M_{di}+\alpha_d b_{di}\right)
\operatorname{Pois}\!\left(m_{di}\mid b_{di}\right)\\
&\times\exp\!\left[-\frac{(\Gamma-\Gamma_0)^2}{2\sigma_\Gamma^2}\right],
\end{split}
\end{equation}
where $\operatorname{Pois}(n\mid\lambda)=\lambda^n e^{-\lambda}/n!$ and the background expectations $b_{di}$ are nuisance parameters, one per energy bin and dataset. This likelihood is equivalent to the W-statistic commonly used for ON/OFF spectra~\cite{Wachter1979}. The last factor in Eq.~\eqref{eq:likelihood} constrains the photon index, which is needed because signal and continuum are partly degenerate in our band: for increasing $\mu$, the fit compensates by steepening the power law, which leaves room for the signal at high energy.

Published hard X-ray indices of the emission around Sgr~A$^\star$ range from $\Gamma\simeq1.5$--$1.6$ for broadband NuSTAR fits that also include thermal components~\cite{Mori2015,Zhang2017}, to $\Gamma\simeq3.0$--$3.2$ above $20\,$keV with INTEGRAL, from a much larger region~\cite{Belanger2006}. We therefore adopt $\Gamma_0=2.4$, the index preferred by the data under the background-only hypothesis ($\Gamma=2.46\pm0.20$ at $\mu=0$), and $\sigma_\Gamma=0.5$. With this constraint, the continuum at the upper limit reaches $\Gamma=3.0$--$3.3$ over the whole mass range, as steep as the steepest measured slope, but not the much steeper continuum, $\Gamma\gtrsim5$, that the fit reaches without a constraint. A wider constraint, $\sigma_\Gamma=1$, which includes all published values within $1\sigma$, is considered to assess the impact of an alternative choice (Appendix~\ref{app:sensitivity_tests}).

To set limits we use the profile likelihood ratio
\begin{equation}
\label{eq:test_statistic}
q(\mu)=-2\ln\frac{\mathcal L\big(\mu,\hat{\hat{\boldsymbol\theta}}(\mu)\big)}
{\mathcal L\big(\hat\mu,\hat{\boldsymbol\theta}\big)},
\end{equation}
where $\boldsymbol\theta=(\Gamma,\{K_d\},\{b_{di}\})$ are the nuisance parameters, $\hat{\hat{\boldsymbol\theta}}(\mu)$ maximizes $\mathcal L$ at fixed $\mu$, and $(\hat\mu,\hat{\boldsymbol\theta})$ is the global maximum. We require $K_d\ge0$ but allow $\hat\mu<0$, as long as all expected counts are non-negative, so that $q$ follows asymptotically a $\chi^2$ distribution with one degree of freedom~\cite{Cowan2011}. We test for the presence of a signal with $q_0=q(0)$ for $\hat\mu>0$, and $q_0=0$ otherwise, whose square root gives the local significance $Z_0=\sqrt{q_0}$~\cite{Cowan2011}. The one-sided $95\%$ C.L.~upper limit $\mu_{95}>\hat\mu$ is the solution of $q(\mu_{95})=2.71$, and $g_{95}=g_0\,\mu_{95}^{1/4}$. We repeat this for axion masses between $10^{-12}$ and $10^{-8}\,$eV, each with its own signal template $\phi_{a,0}(E;m_a)$.

The expected sensitivity is computed from the Asimov dataset~\cite{Cowan2011}, in which the ON and OFF spectra equal the predictions of the best fit with $\mu=0$. The limit $\mu_{95}^{\rm A}$ derived from it is the median limit expected in the absence of a signal. Asymptotically, the standard deviation of $\hat\mu$ is $\sigma_{\rm A}\simeq\mu_{95}^{\rm A}/1.645$, and the band that contains $68\%$ of the expected limits is $\mu_{95}=\sigma_{\rm A}(1.645\pm1)$.

\section{Results}
\label{sec:results}

\begin{figure}
    \centering
    \includegraphics[width=\linewidth]{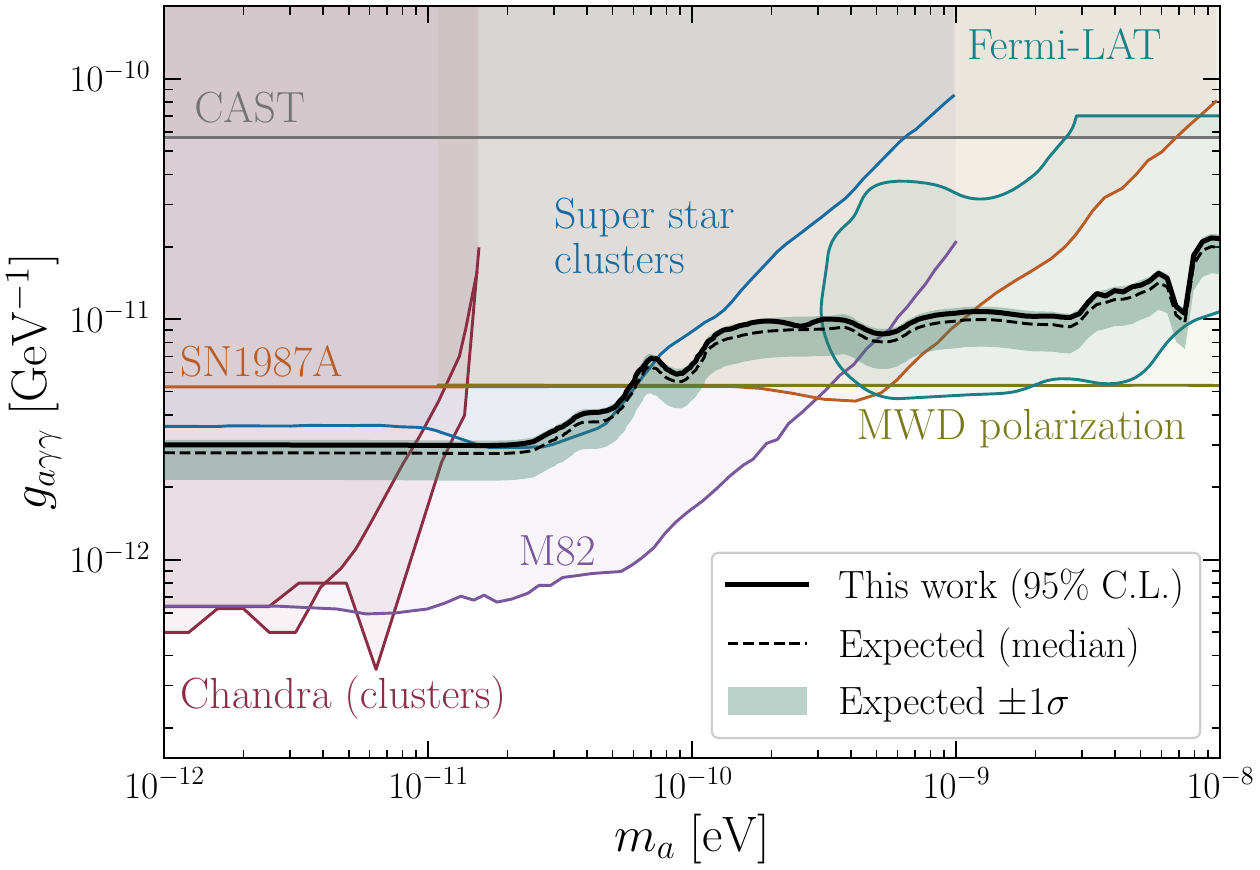}
    \caption{Upper limit at $95\%$ confidence on $g_{a\gamma\gamma}$ from the
    NuSTAR observations of the YNC (black solid line), with the median expected
    limit (dashed) and the $\pm1\sigma$ band of the expected limits (green), for
    the fiducial stellar population and GC model. Colored lines
    and shaded regions show existing bounds: CAST~\cite{CAST2024}, SN1987A
    $\gamma$ rays~\cite{HoofSchulz2023}, super star
    clusters~\cite{DessertClusters2020}, NuSTAR observations of
    M82~\cite{NingSafdi2025}, Chandra spectra of the cluster-hosted AGN
    NGC~1275 and H1821+643~\cite{Reynolds2020,SiskReynes2022}, magnetic white
    dwarf polarization~\cite{DessertMWD2022} and Fermi-LAT observations of
    NGC~1275~\cite{FermiLAT2016}, as provided in Ref.~\cite{AxionLimits}.}
    \label{fig:limits}
\end{figure}

We find no evidence of an axion signal at any mass: the best fit prefers a small positive signal, with local significance $Z_0=0.02$--$1.02$ (median $0.65$), so that even the largest excess would arise from the background alone in $15\%$ of the cases.

Figure~\ref{fig:limits} shows the resulting $95\%$ C.L.~upper limit on $g_{a\gamma\gamma}$, which lies $4$--$11\%$ above the median expected limit and within the $68\%$ band at all masses.

On the plateau below $m_a\simeq2\times10^{-11}\,$eV we obtain $g_{a\gamma\gamma}<3.0\times10^{-12}\,\mathrm{GeV^{-1}}$ (expected $2.8\times10^{-12}\,\mathrm{GeV^{-1}}$), about $19$ times below the CAST limit~\cite{CAST2024}. The limit weakens to $9.9\times10^{-12}\,\mathrm{GeV^{-1}}$ at $3\times10^{-10}\,$eV, $1.3\times10^{-11}\,\mathrm{GeV^{-1}}$ at $4\times10^{-9}\,$eV and $2.2\times10^{-11}\,\mathrm{GeV^{-1}}$ at $10^{-8}\,$eV. Its shape follows that of $\bar P$ in Fig.~\ref{fig:Pbar}, reflecting the importance of the conversion in the GC: the plateau comes from conversion in the inner kiloparsec, the rise above $3\times10^{-11}\,$eV from the loss of coherence on these scales, and the
structure at higher masses from the CMZ, the central cavity and the CND. 
On the other hand, with the JF12 field alone~\cite{JanssonFarrar2012}, the limit would be weaker by a factor $1.8$ at $10^{-12}\,$eV, $5.6$ at $3\times10^{-10}\,$eV and $54$ at $4\times10^{-9}\,$eV.

Figure~\ref{fig:limits} places our limit among the existing bounds in this mass range. The closest analogue is the search for axions from super star clusters~\cite{DessertClusters2020}, which also relies on the Primakoff emission of a young massive population. Our limit is comparable to or stronger than it below $1.5\times10^{-11}\,$eV and above $7\times10^{-11}\,$eV, and it relies on a different source, the YNC, and on a detailed model of the conversion from the central parsecs to the inner kiloparsec. Below $6\times10^{-11}\,$eV, our limit is about $1.7$ times stronger than those from SN1987A~\cite{HoofSchulz2023} and from the polarization of magnetic white dwarfs~\cite{DessertMWD2022}, and above $10^{-9}\,$eV it is also stronger than the SN1987A bound. Only the NuSTAR observations of M82~\cite{NingSafdi2025}, the Chandra spectra of the cluster-hosted AGN NGC~1275 and H1821+643~\cite{Reynolds2020,SiskReynes2022} below $10^{-11}\,$eV and the super star clusters (marginally, by at most $20\%$) are more constraining at low masses. M82 is four to six times more constraining up to $10^{-10}\,$eV and remains stronger up to $5\times10^{-10}\,$eV. At higher masses, the white dwarfs give limits up to four times stronger, and SN1987A and the Fermi-LAT observations of NGC~1275~\cite{FermiLAT2016} up to $2.2$ and $2.4$ times stronger, respectively.  These searches rely on different sources, magnetic-field models and instruments, so that our limit, based on a Galactic source with a resolved stellar population and an explicit model of the conversion, provides an independent test of the same parameter space.

\begin{figure}
    \centering
    \includegraphics[width=\linewidth]{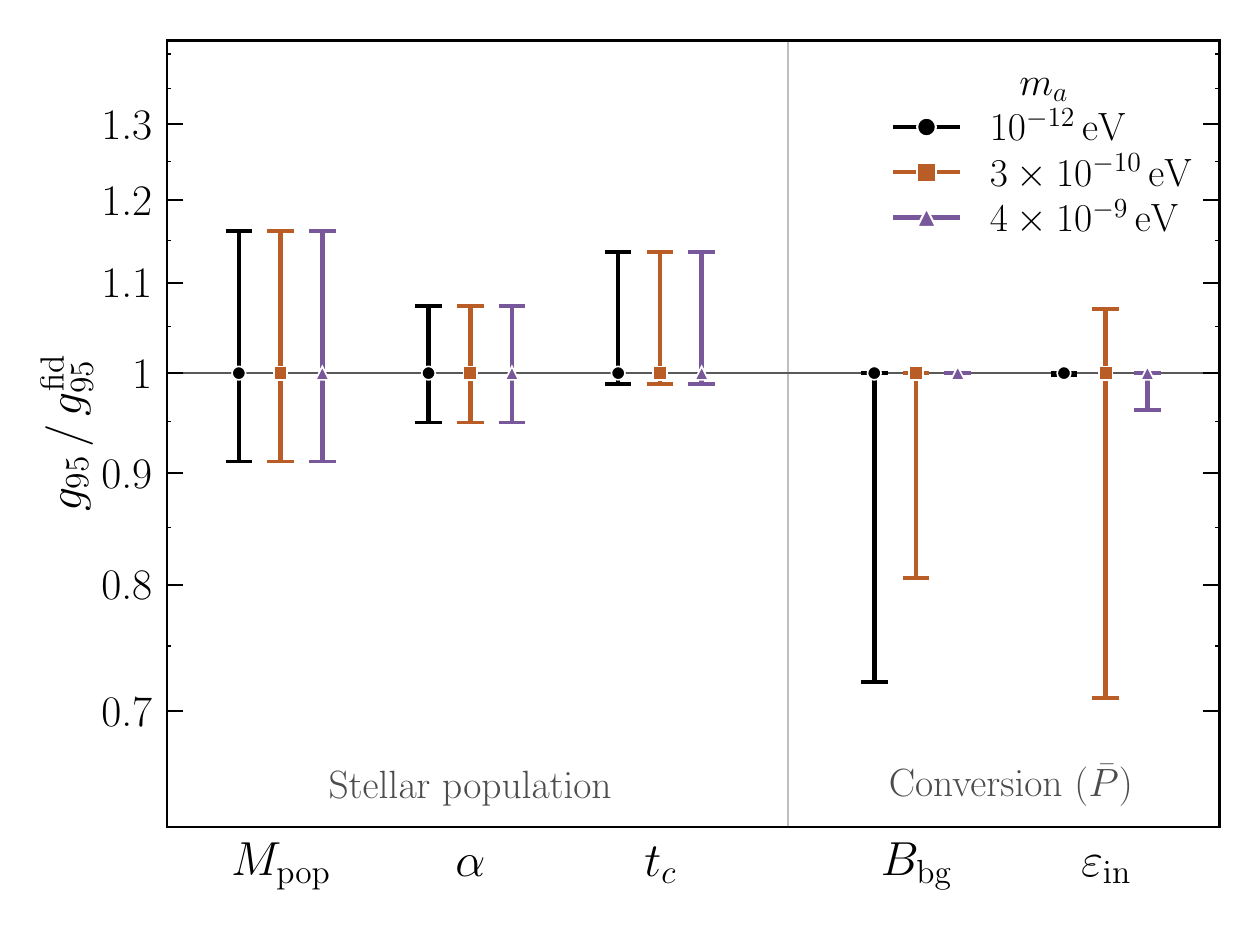}
    \caption{Ratio of the $95\%$ C.L.~upper limit to the fiducial one, $g_{95}/g_{95}^{\rm fid}$, when each input is varied over its observational interval, one at a time, at three benchmark masses (colors): the cluster mass $M_{\rm pop}$, the IMF slope $\alpha$, and the $68\%$ interval of the central age $t_c$ from Ref.~\cite{Lu2013IMF}; the background field $B_{\rm bg}$, varied from the fiducial $10\,\mu$G to $50\,\mu$G within $400\,$pc~\cite{Crocker2010}; and the floor $\varepsilon_{\rm in}$ of the CND field in the central cavity, varied from $0.15$ to $0.38$, corresponding to the measured median plane-of-sky field of $2$--$5\,$mG in the minispiral~\cite{Akshaya_2024}.}
    \label{fig:systematics}
\end{figure}

\paragraph*{Systematic uncertainties.}
Figure~\ref{fig:systematics} shows how the limit changes when the inputs of the signal model with an observational uncertainty are varied over their quoted intervals, one at a time, at three masses representative of the low-mass plateau ($10^{-12}\,$eV), of the conversion in the CMZ and in the central cavity ($3\times10^{-10}\,$eV), and of the resonances in the central parsecs ($4\times10^{-9}\,$eV). The stellar population inputs of Ref.~\cite{Lu2013IMF} change the limit by the same factor at all masses, since they mainly set the normalization of the spectrum: by $-9\%$/$+16\%$ over the range of the initial cluster mass, by $-5\%$/$+7\%$ for the IMF slope, and by $-1\%$/$+14\%$ over the $68\%$ interval of the central age. The magnetic field represents the main source of uncertainty: adopting the lower limit of $50\,\mu$G within $400\,$pc of Ref.~\cite{Crocker2010} instead of our $10\,\mu$G strengthens the limit by $28\%$ on the plateau and by $19\%$ at $3\times10^{-10}\,$eV, while varying $\varepsilon_{\rm in}$ over $0.15$--$0.38$, which gives the measured $2$--$5\,$mG in the minispiral~\cite{Akshaya_2024}, changes the limit by $-29\%$/$+7\%$ around $3\times10^{-10}\,$eV and by less than $5\%$ elsewhere. Our fiducial values thus represent conservative choices in both intervals.
We refer to Appendix~\ref{app:sensitivity_tests} for a discussion of sensitivity tests, in which modeling assumptions and choices of the X-ray analysis are replaced by alternatives.

\section{Discussion and conclusions}
\label{sec:conclusions}

We have searched for axions produced by the young stellar population within the central $0.5\,$pc of the Milky Way nuclear star cluster, the YNC, and converted into hard X rays in the Galactic magnetic field. We combined stellar evolution tracks with the population parameters inferred by Lu et al.~\cite{Lu2013IMF} to model the YNC, whose axion emission is dominated by stars of $60$--$100\,M_\odot$ in advanced evolutionary phases. We described the conversion along the whole path of the axions, from the central parsecs to the Galactic disk. Our description of the inner Galaxy builds on the gas and magnetic-field models of Refs.~\cite{ferriere2007,Ferriere2012,Guenduez2016}, with the modifications of Sec.~\ref{sec:galactic_environment}. The NuSTAR observations show no evidence for an axion signal and constrain $g_{a\gamma\gamma}<3.0\times10^{-12}\,\mathrm{GeV^{-1}}$ at $95\%$ C.L.~for $m_a\lesssim2\times10^{-11}\,$eV. At higher masses, where the conversion takes place in the CMZ and in the central parsecs, the limit weakens to about $10^{-11}\,\mathrm{GeV^{-1}}$ around the neV scale and to $2\times10^{-11}\,\mathrm{GeV^{-1}}$ at $10^{-8}\,$eV.

To our knowledge, this is the first search focused on this target. The closest target and type of analysis to the one performed here is the search toward super star clusters reported in Ref.~\cite{DessertClusters2020}, whose sensitivity we match or exceed at all masses except $m_a\simeq(1.5$--$7)\times10^{-11}\,$eV, where their bound is stronger by at most $20\%$. Further, we extend the bound to higher masses. While slightly weaker, our bounds remain competitive with other limits from X-ray observations of AGN and of M82 below $\sim10^{-10}\,$eV and from the polarization of magnetic white dwarfs at higher masses. Our limit is stronger than the SN1987A and white dwarf bounds below $6\times10^{-11}\,$eV. At higher masses it is at most $2.4$ times weaker than the SN1987A and Fermi-LAT bounds, and stronger than the SN1987A bound above $10^{-9}\,$eV.

We emphasize that multiplying the astrophysical axion studies is valuable because various searches rely on different sources, environments, and detectors. The inputs of each search are affected by uncertainties of various kinds, with more exotic objects (like compact stars) and farther systems (as objects in external galaxies) generally subject to bigger systematics. 

In our fiducial model, the inner Galaxy provides most of the conversion throughout the mass range studied: the adopted coherent field in the inner kiloparsec controls the low-mass plateau, while the CMZ and the central parsecs dominate at higher masses (Sec.~\ref{sec:Pbar} and Fig.~\ref{fig:Pbar}). For the YNC, restricting the calculation to the large-scale JF12 field~\cite{JanssonFarrar2012} therefore substantially underestimates the predicted signal. This motivates an explicit assessment of the inner-Galaxy contribution in other searches toward the GC, whose sensitivity to these structures depends on the source position, the photon energy and the axion mass. 
At hard X-ray energies, we stress that the effective photon mass must also include bound electrons, which dominate in molecular gas, as implemented in this work. To get a sense of the error incurred by neglecting it in this type of searches,  we computed it from the free electrons alone, keeping our gas model. In this case, the limit would be stronger by up to a factor of $2$ between $10^{-10}$ and $10^{-9}\,$eV, but weaker by up to a factor of $5$ above $4\times10^{-9}\,$eV. Since the size and even the sign of this effect depend on the gas and magnetic fields crossed by each line of sight, we emphasize it has to be assessed for each target, be it Galactic or extragalactic.

The main source of systematic uncertainty of the present study is the magnetic field of the inner Galaxy. At low masses, our fiducial background field of $10\,\mu$G is conservative; implementing in the inner 400 pc even the lower limit reported in Ref.~\cite{Crocker2010}, of $50\,\mu$G, would strengthen the limit by a factor of $1.4$. The stellar population and the analysis choices change the limit by less than $\sim 20\%$.  A better knowledge of the strength and geometry of the magnetic field in the inner Galaxy, for instance from dust polarimetry~\cite{Guerra_2023} and Faraday rotation measurements in the CMZ, would reduce the uncertainty in the predicted signal and make the inferred constraint more robust. We have also neglected turbulent magnetic fields, which may increase the conversion, particularly at high masses where the coherent field converts inefficiently, as shown for supernova axions in Ref.~\cite{Fiorillo:2026rgo}. Neglecting them is therefore conservative. We also neglect the conversion in the magnetic fields of the stars themselves. In our population model, most of the axion emission comes from compact stars with $R_\star\sim2\,R_\odot$. For each of them, even a kG dipole, a generous assumption~\cite{delaChevrotiere2013}, gives a conversion probability about three orders of magnitude below $\bar P$ at all masses.

We note that additional exposure would improve the statistical sensitivity, but only slowly: as long as the sensitivity is limited by the statistical uncertainty of a dominant background, the limit on $g_{a\gamma\gamma}$ scales as $t^{-1/8}$ with the exposure time $t$. 

Finally, our state-of-the-art analysis of axion-photon conversion
including recent observational constraints on magnetic field and gas density in the inner parsecs enables us to extend this study to other stellar populations of the GC, such as the Arches and Quintuplet clusters and the old nuclear star cluster, with their respective source distributions and lines of sight, and for future hard X-ray observations.

\begin{acknowledgments}
We acknowledge M.~Clavel (IPAG) and S.~Simón-Díaz (IAC) for useful discussions on the X-ray analysis and massive stars in the Galaxy, respectively. RZ thanks D.~Grossmann (IAC) for assistance with the MESA stellar evolution code.
This work is supported by the European Union through the grant ``UNDARK'' of the Widening participation and spreading excellence programme (project number 101159929). RZ and JMC also acknowledge the MICINN through the grant ``DarkMaps'' PID2022-142142NB-I00 and the grant CEX2025-001609-S, awarded to the Instituto de Astrofísica de Canarias under the Severo Ochoa Centre of Excellence program and funded by MICIU/AEI/10.13039/501100011033.
The authors used AI tools (Claude Opus 5.5 and GPT 6.2) to assist with drafting analysis scripts, organizing literature notes, and improving the grammar and readability of the text. All code, references, and outputs were verified and validated against known results by the authors.
 \end{acknowledgments}

\clearpage
 
\onecolumngrid
\appendix

\section{Stellar models and population spectrum}
\label{app:stellar_models}

\subsection{Stellar models and single-star spectra}
\label{app:mesa_evolution}

We compute the stellar profiles with MESA, release \texttt{r24.08.1}~\cite{Jermyn2023MESA}, using a setup based on the
massive star suite \texttt{20M\_pre\_ms\_to\_core\_collapse}. We fix the initial composition to $Y=0.28$ and $Z=0.02$, with
the relative metal abundances of Ref.~\cite{Asplund2009}, and the initial surface rotation to $300\,$km\,s$^{-1}$ near the zero-age main sequence (ZAMS). The metallicity and initial rotation match those of rotating massive star models computed with the Geneva stellar evolution code~\cite{Eggenberger2008}, which Ref.~\cite{Lu2013IMF} used to model the young stars.
The remaining input physics follows the MESA test case: MLT++, which treats the superadiabatic envelopes of massive stars, is active, and mass loss follows the \texttt{Dutch} scheme.

We evolve the stars from the ZAMS to central carbon depletion, $X_{{\rm C12},c}<10^{-3}$, and do not follow the subsequent evolution to core collapse. In our models, core carbon burning lasts about 10 yr for $60$--$100\,M_\odot$, and the later burning stages last at most of order a year~\cite{Woosley2002}.

Along each track, we compute the axion spectrum from the stellar profiles saved by MESA, with finer sampling after central helium depletion, when the stellar structure evolves faster. Each profile provides the temperature, density and composition of every cell; the composition gives the number densities of free electrons and ions that enter $\kappa_s$ and $\omega_{\rm pl}$ in Eq.~\eqref{eq:Primakoff_emissivity}. The axion luminosity of Eq.~\eqref{eq:Luminosity_Primakoff} is then the sum of the emissivity over the cells.

\begin{figure*}[t]
    \centering
    \includegraphics[width=\textwidth]{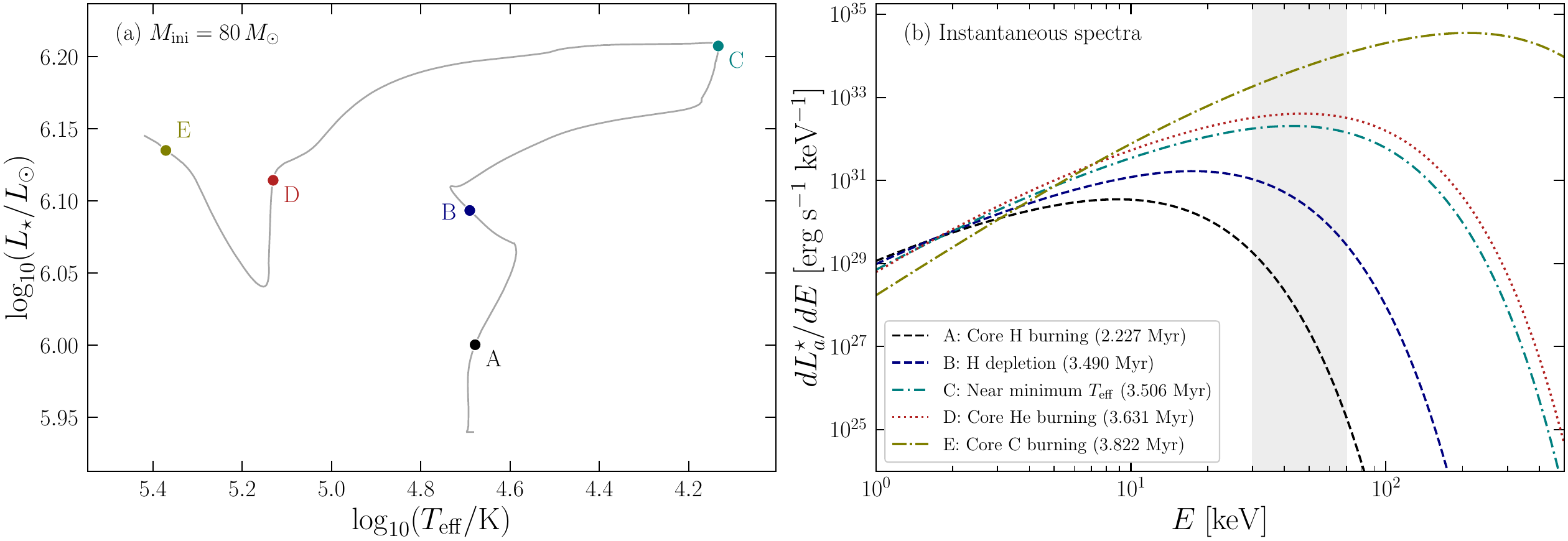}
    \caption{Evolution and instantaneous Primakoff spectra of a star with
    initial mass $80\,M_\odot$, for
    $\gagg=10^{-12}\,\mathrm{GeV}^{-1}$.
    Left: Hertzsprung--Russell diagram from the MESA history, with five
    saved profiles marked along the track. Right: spectra computed from
    those profiles, with the same colors and labels.
    The profiles sample core hydrogen burning (A), the end of the main
    sequence (B), the minimum effective temperature (C), core helium
    burning (D) and core carbon burning (E).
    The legend reports the ages of the selected profiles.
    The shaded band indicates $30$--$70\,$keV.}
    \label{fig:mesa_single_star_stages}
\end{figure*}

Figure~\ref{fig:mesa_single_star_stages} illustrates the relation between stellar evolution and axion emission for a star of initial mass $80\,M_\odot$, representative of the stars that dominate the emission of the population (see Appendix~\ref{app:population_spectrum}). We select five evolutionary stages from the central abundances and effective temperature recorded in the MESA history, and compute the associated spectrum. 
The spectra become harder as the stellar interior heats up, since the emission follows the thermal photon distribution, the factor $(e^{E/T}-1)^{-1}$ in Eq.~\eqref{eq:Primakoff_emissivity}, and peaks at a few times the local temperature; the energy-integrated emissivity also grows rapidly with temperature, approximately as $T^7$~\cite{Raffelt1986}. In the $30$--$70\,$keV band, the axion luminosity of a single star increases by about three orders of magnitude between the end of the main sequence and central helium depletion, where it reaches a few $\times10^{34}\,$erg\,s$^{-1}$ for initial masses $\gtrsim40\,M_\odot$.

We finally note that the spectra in Fig.~\ref{fig:mesa_single_star_stages} are instantaneous and do not represent the weight of each stage in the population spectrum, which also depends on the stage duration and on the distribution of stellar ages (Appendix~\ref{app:population_spectrum}).

\subsection{Population spectrum}
\label{app:population_spectrum}

Next, we account for the finite duration of the star formation episode by
describing the stellar ages with a uniform distribution centered on $t_c$,
of full width $\Delta t$, and by averaging the spectrum of each star over it.
For a star born with initial mass $M$, the burst-averaged differential
axion luminosity at energy $E$ is
\begin{equation}
    \label{eq:burst_average_spectrum}
    \left\langle\frac{\rd L_a^\star}{\rd E}\right\rangle(E;M,t_c,\Delta t)
    =\frac{1}{\Delta t}
    \int_{t_c-\Delta t/2}^{t_c+\Delta t/2}
    \rd t\,\frac{\rd L_a^\star}{\rd E}(E;M,t)\,
    \Theta\!\left[t_{\rm end}(M)-t\right].
\end{equation}
Here $t_{\rm end}$ is the age of the last saved profile, after which we
set the contribution to zero. Between consecutive profiles, we
interpolate the logarithm of the spectrum linearly in age. The denominator remains the full
$\Delta t$ even when a track ends within the window: renormalizing to the
surviving interval would overcount the contribution of short-lived stars.

The initial masses follow the IMF $\xi(M)\propto M^{-\alpha}$, normalized to the initial population mass $M_{\rm pop}$ over
$1\leq M/M_\odot\leq150$, the range adopted in Ref.~\cite{Lu2013IMF};
we use the fiducial values $\alpha=1.7$, $M_{\rm pop}=2.55\times10^4\,M_\odot$,
$t_c=3.9\,$Myr and $\Delta t=1\,$Myr (as described in Sec.~\ref{sec:population_axion_spectrum}).
We compute the stellar models for 32 initial masses between $20$ and $150\,M_\odot$.
Each grid mass represents a bin bounded by the midpoints between
neighboring masses, with outer boundaries at $20$ and $150\,M_\odot$. We compute the population spectrum by summing the burst-averaged spectra of each mass bin, weighted by the expected number of stars in that bin:
\begin{equation}
    \label{eq:population_spectrum}
    \left\langle\frac{\rd L_a}{\rd E}\right\rangle(E)=\sum_k N_k
    \left\langle\frac{\rd L_a^\star}{\rd E}\right\rangle(E;M_k,t_c,\Delta t),
    \qquad
    N_k=\int_{{\rm bin}\,k}\xi(M)\,\rd M,
\end{equation}
where $N_k$ is the expected number of stars in the bin of the grid mass $M_k$.

As discussed in Appendix~\ref{app:mesa_evolution}, the luminosity of massive stars rises steeply after the main sequence, and this is reflected in the population spectrum. For the fiducial parameters, the population signal is dominated by the stars that are past the main sequence during the burst window, $3.4$--$4.4\,$Myr: the mass bins between $60$ and $100\,M_\odot$
contribute about $95\%$ of the $30$--$70\,$keV luminosity. Stars with $M\leq50\,M_\odot$ remain on the main sequence throughout the window and contribute about $10^{-5}$, while stars with $M\gtrsim120\,M_\odot$ reach the end of their computed tracks before the window begins. For this reason, we build a finer grid in the dominant range, with a spacing of $2.5\,M_\odot$ between $60$ and $100\,M_\odot$; we test its convergence in Appendix~\ref{app:burst_average}. Stars born below $20\,M_\odot$, which are even less evolved, are not included in the axion luminosity calculation.

\subsection{Dependence on the population parameters}
\label{app:burst_average}
\begin{figure*}[t]
    \centering
    \includegraphics[width=\textwidth]{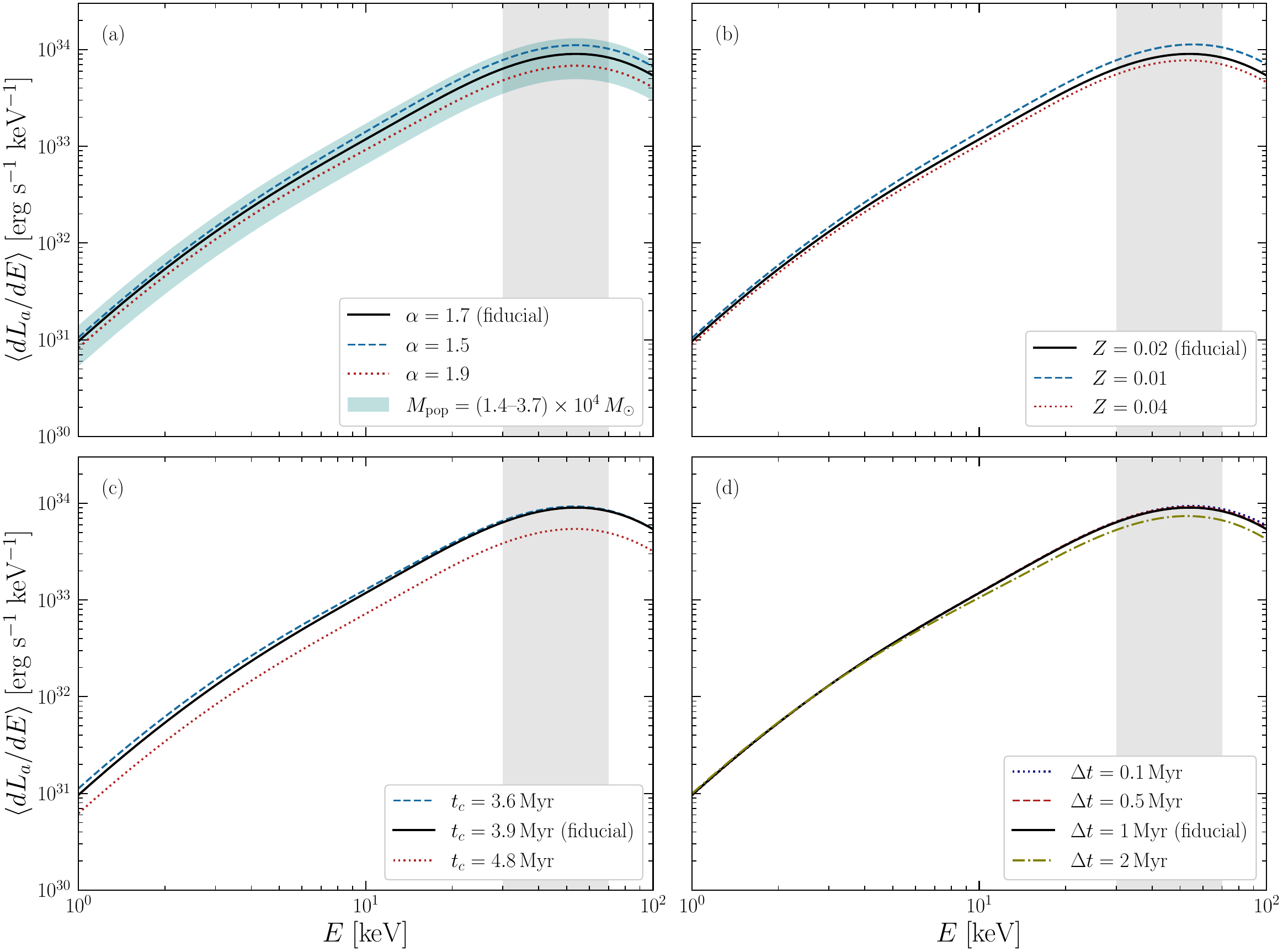}
    \caption{Population spectrum of the YNC for
    $\gagg=10^{-12}\,\mathrm{GeV}^{-1}$ and different population parameters,
    each varied with the others fixed at their fiducial values
    ($\alpha=1.7$, $M_{\rm pop}=2.55\times10^4\,M_\odot$, $t_c=3.9\,$Myr,
    $\Delta t=1\,$Myr, $Z=0.02$). (a)~IMF slope $\alpha=1.5$ and $1.9$; the teal band spans
    $M_{\rm pop}=(1.4$--$3.7)\times10^4\,M_\odot$~\cite{Lu2013IMF}.
    (b)~Initial metallicity $Z=0.01$ and $0.04$. (c)~Central age $t_c=3.6$
    and $4.8\,$Myr, the $68\%$ interval of the main age solution of
    Ref.~\cite{Lu2013IMF}. (d)~Top-hat
    age distributions of full width $\Delta t=0.1$, $0.5$, $1$ and $2\,$Myr.
    The gray shaded band indicates $30$--$70\,$keV.}
    \label{fig:population_variants}
\end{figure*}

We deal here with the uncertainties in the population parameters (IMF slope, total mass, central age and duration of star formation, and in the initial metallicity) by varying them one at a time, while keeping the others fixed at their fiducial values. The effect on the limit is given in Sec.~\ref{sec:results} and Appendix~\ref{app:sensitivity_tests}. Figure~\ref{fig:population_variants}(a) shows the population spectrum for the IMF slopes $\alpha=1.5$ and $1.9$, the $1\sigma$ range of Ref.~\cite{Lu2013IMF}, and for the range of initial population mass $M_{\rm pop}=(1.4$--$3.7)\times10^4\,M_\odot$ inferred there. In practice, both these modifications amount to a rescaling of the spectrum.

Figure~\ref{fig:population_variants}(b) shows the effect of the initial metallicity. Observations of the young stars in the GC indicate an iron abundance consistent with solar, possibly up to twice solar (see the
discussion in Ref.~\cite{Lu2013IMF}); we consider $Z=0.04$ and, as a symmetric test, $Z=0.01$. For both values, we recomputed the tracks of the 24-mass grid, with $Y=0.28$ and otherwise the same setup. For the fiducial burst, this grid reproduces the $30$--$70\,$keV luminosity of the 32-mass one within $1.1\%$ (see below), which is a negligible difference for our purposes. Halving $Z$ increases the $30$--$70\,$keV luminosity by about $25\%$, and doubling it reduces the luminosity by about $14\%$, with ratios that depend only mildly on energy within the band.

Figure~\ref{fig:population_variants}(c) shows the effect of the central age. The age posterior of Ref.~\cite{Lu2013IMF} is bimodal: its main solution contains the mode at $3.9\,$Myr, which we adopt, and a secondary one peaks around $2.8\,$Myr. Spectroscopic studies of the Wolf--Rayet stars and OB supergiants favor ages $\gtrsim4\,$Myr~\cite{Paumard:2006}, as does the best fit of Ref.~\cite{Gallego-Cano}. We therefore vary $t_c$ within the $68\%$ interval of the main solution, $3.6$--$4.8\,$Myr, keeping $\Delta t=1\,$Myr. The $30$--$70\,$keV luminosity reaches a maximum about $5\%$ above the fiducial value at $3.7\,$Myr and is $40\%$ lower at $4.8\,$Myr: at later ages, the most massive stars have already ended their evolution, and the band is dominated by less massive, fainter stars. 

We also compare finite burst widths $\Delta t=0.1$, $0.5$, $1$ and $2\,$Myr, centered at $t_c=3.9\,$Myr. The widest window, $2.9$--$4.9\,$Myr, corresponds to the maximal duration of the formation episode suggested by Ref.~\cite{Paumard:2006}; the narrowest, $3.85$--$3.95\,$Myr, approximates a coeval population. Such a narrow burst requires particular care with the mass sampling, since brief luminous phases occur at different ages in adjacent tracks.

Figure~\ref{fig:population_variants}(d) isolates the effect of the burst width by fixing $t_c=3.9\,$Myr, $\alpha=1.7$ and $M_{\rm pop}=2.55\times10^4\,M_\odot$. Changing $\Delta t$ changes which evolutionary phases of each star fall within the age window, and with what fraction. Increasing the width from $1$ to $2\,$Myr reduces the $30$--$70\,$keV luminosity by about $18\%$, while reducing it (i.e.~approaching the coeval limit) always increases the luminosity, by about $1.7\%$ for $\Delta t=0.5\,$Myr and $3.8\%$ for $\Delta t=0.1\,$Myr.

We assess the mass sampling by comparing grids of 24 and 32 masses, with spacings of $5$ and $2.5\,M_\odot$ between $60$ and $100\,M_\odot$. Between the two, the $30$--$70\,$keV luminosity changes by at most $1.1\%$ for $\Delta t=0.5$, $1$ and $2\,$Myr, and by $4.9\%$ for $\Delta t=0.1\,$Myr, comparable to the difference from the fiducial burst. The $3.8\%$ enhancement of the $0.1\,$Myr burst is therefore within the sampling uncertainty.

Overall, the population parameters affect mainly the normalization of the spectrum: together, the cluster mass and the IMF slope change the $30$--$70\,$keV luminosity by about a factor of two (panel a), and the central age by up to $40\%$ (panel c). Since the limit scales as $L_a^{-1/4}$, the corresponding changes of $g_{a\gamma\gamma}$ are much smaller (see Sec.~\ref{sec:results} of the main text).

\section{Galactic Center environment models}
\label{app:galactic_center_models}

This appendix details the gas-density and magnetic-field models adopted to describe the GC environment, including their geometries and the modifications introduced in our implementation.

We use Galactocentric Cartesian coordinates $(x,y,z)$, with Sgr~A$^\star$ at the origin, the $x$ axis pointing toward the Sun, the $y$ axis toward positive Galactic longitude, and the $z$ axis toward the north Galactic pole. We denote the cylindrical Galactocentric radius by $r=\sqrt{x^2+y^2}$. We adopt a GC distance of $D_{\rm GC}=8.178\,$kpc~\cite{GRAVITY2019} and place the observer at $(D_{\rm GC},0,0)$.

\subsection{Gas and electron-density components}
\label{app:density_components}

\begin{table*}[t]
\caption{
Adopted gas components in the central Galactic region. Centers and dimensions refer to the Galactocentric frame, except for the CND, which is specified in its inclined frame. Densities refer to hydrogen nuclei.
}
\label{tab:inner_gc_gas}
\centering
\begingroup
\small
\renewcommand{\arraystretch}{1.18}
\setlength{\tabcolsep}{5pt}

\begin{tabular*}{\textwidth}{
    @{\extracolsep{\fill}} l c l r @{}
}
\toprule
\textbf{Component}
& \textbf{Center}
& \textbf{Geometry}
& \boldmath$n_{\rm H}$ \\
& $(x_c,y_c,z_c)$ [pc]
& Dimensions in pc
& [cm$^{-3}$] \\
\midrule

Radio halo
& $(-2.0,\,1.2,\,-1.5)$
& Sphere, $R=9.0$
& $210$ \\

\addlinespace[3pt]
20 km\,s$^{-1}$ cloud (SC)
& $(8.0,\,-11.0,\,-5.0)$
& Ellipsoid, $(a,b,c)=(3.75,\,7.5,\,3.75)$
& $2.0\times10^4$ \\

\addlinespace[3pt]
Sgr~A~East SNR
& $(-2.0,\,1.2,\,-1.5)$
& Ellipsoid, $(a,b,c)=(4.5,\,4.5,\,3.35)$
& $3$ \\

\addlinespace[3pt]
Central cavity (CC)\textsuperscript{a}
& $(0,\,0,\,0)$
& Ellipsoid, $(a,b,c)=(1.45,\,1.45,\,1.05)$
& $910$ \\

\addlinespace[3pt]
50 km\,s$^{-1}$ cloud (EC)
& $(-3.0,\,7,\,-4.5)$
& Sphere, $R=4.5$
& $2.0\times10^4$ \\

\addlinespace[3pt]
CND: inner layer\textsuperscript{b}
& $(0,\,0,\,0)$
& Annulus, $1.2< r'\lesssim1.6$
& $3.2\times10^4$ \\

\quad Outer layer
& $(0,\,0,\,0)$
& Annulus, $1.6\lesssim r'<3.0$
& $4.4\times10^5$ \\

\bottomrule
\end{tabular*}

\par\vspace{5pt}
\begin{minipage}{\textwidth}
\footnotesize
\raggedright

Geometries approximate the structures described in
Refs.~\cite{Ferriere2012,Guenduez2016};
$(a,b,c)$ denote semiaxes.
Dimensions are nominal values before smoothing.
\par
\textsuperscript{a}
The cavity also contains the schematic minispiral described below.
\par
\textsuperscript{b}
The CND full thickness increases from $0.4$ to $1.0\,$pc
between its inner and outer edges; see text for its orientation and
smooth boundaries. 
\end{minipage}

\endgroup
\end{table*}

All gas densities count hydrogen nuclei, $n_{\rm H}$; the corresponding
electron density is $n_e=1.2\,n_{\rm H}$, as discussed in
Sec.~\ref{sec:gas_plasma}.
We denote by $n_{\rm F07}$ and $n_{\rm F98}$
the hydrogen densities obtained by combining the
molecular, atomic, and ionized components of
Refs.~\cite{ferriere2007} and \cite{Ferriere1998}, respectively:
\begin{align}
    n_{\rm F07}&=2n_{\rm H_2}+n_{\rm HI}+n_{\rm H^+},
    \label{app:n_F07}\\
    n_{\rm F98}&=n_{\rm m}+n_{\rm c}+n_{\rm w}+n_{\rm i}+n_{\rm h}.
    \label{app:n_F98}
\end{align}
Equation~\eqref{app:n_F07} is Eq.~(29) of Ref.~\cite{ferriere2007}: the
molecular and atomic terms include the CMZ and the tilted, holed bulge
disk, and the ionized term the warm, hot and very hot ionized media
(their Eqs.~30--32). For the CMZ, we use the renormalized expressions of
Eqs.~(1)--(2) of Ref.~\cite{Guenduez2016} (hereafter G20). Equation~\eqref{app:n_F98} sums
the hydrogen densities of the molecular gas, the cold and warm
neutral media, the warm ionized medium and the hot medium of
Ref.~\cite{Ferriere1998}, as in their Eq.~(14); the individual terms are
their Eqs.~(2), (6), (7), (9) and (13).
We use $n_{\rm F07}$ for $r\leq3\,$kpc and $n_{\rm F98}$ for
$r\geq3.5\,$kpc, and join them as
$n_{\rm bg}=(1-S)\,n_{\rm F07}+S\,n_{\rm F98}$, where $S$ is evaluated at
$u=(r-3\,{\rm kpc})/(0.5\,{\rm kpc})$ and is the smoothstep function
\begin{equation} \label{app:smoothstep}
    S(u)=\begin{cases}0,&u\le0,\\3u^2-2u^3,&0<u<1,\\1,&u\ge1.\end{cases}
\end{equation}
Starting from $n_{\rm bg}$, we insert the local structures of
Ref.~\cite{Ferriere2012} with the geometries and densities in
Table~\ref{tab:inner_gc_gas}, using
\begin{equation}\label{app:replacement}
    n_{\rm new}= (1-G_A)n_{\rm old}+G_An_A,
\end{equation}
where $n_A$ is the density profile assigned to the structure and $G_A$ is its spatial gate,
which decreases from one inside it to zero outside. We apply the
structures in table order, so that later entries take precedence in
overlapping regions, such as the CND and CC.

Smooth transitions avoid density discontinuities and provide finite
widths for the confined magnetic-field boundaries
(Appendix~\ref{app:magnetic_field}). For the radio halo, SNR, CC and the MCs, we
define the normalized distance $q$ from the center of the structure,
$q=|\mathbf x-\mathbf x_A|/R$ for a sphere of radius $R$ and
$q^2=\sum_i(\Delta x_i/a_i)^2$ for an ellipsoid of semiaxes $a_i$, and use
the logistic gate
\begin{equation}\label{app:cloud_gate}
    G_A=\left[1+e^{(q-1)/\epsilon}\right]^{-1},\qquad
    \epsilon=\frac{0.10}{2\ln9}\simeq0.023,
\end{equation}
whose $90$--$10\%$ transition width is $10\%$ of the radius or of the
directional semiaxis. Besides SC and EC, we include the other MCs
of G20 (Sgr~C, the Dust Ridge clouds,
Sgr~B2 and Sgr~D), with the positions, radii and densities given there;
they lie tens of parsecs or more from Sgr~A$^\star$ and do not intersect the
lines of sight considered here.

\subsubsection*{The CND}

We adopt the CND orientation described in
Ref.~\cite{Ferriere2012}: its plane is inclined by $70^\circ$
to the plane of the sky, with the line of nodes at a position
angle of $25^\circ$ east of equatorial north.
In our Galactocentric coordinate system, this corresponds
to the unit normal
$\hat{\mathbf n}_{\rm CND}\simeq(0.342,\,0.105,\,-0.934)$.
We denote coordinates in the disk frame by $(x',y',z')$,
with $z'$ along $\hat{\mathbf n}_{\rm CND}$, and define
$r'=\sqrt{x'^2+y'^2}$.

The CND radial gate combines its inner and outer boundaries,
\begin{equation} \label{app:CND_radial_gate}
    G_{{\rm CND},r}(r')=
    \frac{1}{1+e^{(r'_{\rm in}-r')/w_r}}
    \frac{1}{1+e^{(r'-r'_{\rm out})/w_r}}, \qquad w_r=0.04\,\rm{pc}.
\end{equation}
Here $r'_{\rm in}=1.2\,$pc, $r'_{\rm out}=3.0\,$pc,
and each edge has a $90$--$10\%$ transition width of approximately
$0.18\,$pc, or $10\%$ of the annular width. The local half-height $h(r')$
increases smoothly from $0.2$ to $0.5\,$pc between the inner and outer
edges, following Eq.~\eqref{app:smoothstep}. We define the distance
outside the disk surface as $d=\max\!\left(|z'|-h(r'),\,0\right)$
and write the vertical gate as
\begin{equation} \label{app:CND_vertical_gate}
    G_{{\rm CND},z}(r',z')=\operatorname{sech}^{2}\!\left(\frac{d}{w_zh(r')}\right), \qquad w_z = 0.134.
\end{equation}
The vertical $90$--$10\%$ transition width is $10\%$ of the local
full thickness $H=2h$. The product
\begin{equation} \label{app:CND_total_gate}
    G_{\rm CND}=G_{{\rm CND},r}\,G_{{\rm CND},z},
\end{equation}
enters the replacement rule of Eq.~\eqref{app:replacement}.

\subsubsection*{Other structures}

Within the central cavity, we include the minispiral (Sgr~A West), a set of ionized gas streamers, with $n_{\rm H}=6200\,{\rm cm^{-3}}$. Since Ref.~\cite{Ferriere2012} provides only the total length ($\approx9.7\,$pc) and diameter ($\approx0.1\,$pc) of its filaments, we model them schematically as three tubes close to the CND plane, within $r'<0.8\,$pc and $|z'|<0.2\,$pc in the CND frame, i.e.\ inside the CND inner edge, to avoid overlaps with the CND gas distribution.

We do not include the localized molecular structures referred to as the Molecular Ridge, Southern Streamer, Western Streamer,
and Northern Ridge in Ref.~\cite{Ferriere2012}. In the adopted geometry, these structures do not intersect the lines of sight considered here. Moreover, no dedicated magnetic-field components are assigned to them in G20.

\subsection{Magnetic-field model}
\label{app:magnetic_field}

We start from the central $\sim200\,$pc, for which we adopt the model of
G20. It describes the central magnetic environment as
the combination of the ICM field, the fields of the MCs and NTFs, and the contribution due to Sgr~A$^\star$:
\begin{equation}\label{app:B_GC}
    \mathbf B_{\rm GC}
    =\mathbf B_{\rm ICM}
    +\mathbf B_{\rm CND}
    +\mathbf B_{\rm MC}
    +\mathbf B_{\rm NTF}
    +\mathbf B_{\rm SgrA^\star}.
\end{equation}
We keep the ICM, NTF and Sgr~A$^\star$ fields of G20 nearly unchanged,
revise the confinement and normalization of the CND and MC fields, and
connect the central model to JF12 through the inner kiloparsec.

\subsubsection*{ICM, NTF and \texorpdfstring{Sgr~A$^\star$}{Sgr A*} fields}

We retain the G20 poloidal ICM and NTF fields. The ICM field has radial scale length $L=114\,$pc, field-line opening parameter $a=(42\,\mathrm{pc})^{-2}$ and normalization $B_1=12\,\mu$G. The NTFs are recentered versions of the same construction, with the positions, dimensions and field strengths listed in Table~3 of G20. The azimuthal field of Sgr~A$^\star$ follows Eqs.~(44)--(45) of G20, normalized to $B_\phi(2.4\,\mathrm{pc},0)=10\,\mu$G.

\subsubsection*{CND and MC fields}

In G20, the CND field (their ``inner $5\,$pc'' component) is normalized
to the equipartition estimate $B_{\rm eq}=8.4\,$mG (their Table~4),
consistent with the several mG, locally reaching tens of mG, measured in
the CND~\cite{Guerra_2023}. Its $1/r$ profile, however, is not confined to
the CND gas: it exceeds $100\,$mG at subparsec radii and is still
$\simeq10\,$mG at $4$--$5\,$pc, whereas observations indicate a median of
2--5 mG in the minispiral of the CC~\cite{Akshaya_2024} and fields of order
$10\,\mu$G in the extended diffuse medium~\cite{LaRosa_2005}. We therefore
keep the form of the G20 horizontal field and its reference strength, but
confine the field to the gas geometry of the CND and renormalize it so that
its average over that gas equals $B_{\rm eq}$; we apply the same
procedure to the other MCs.

We emphasize that this revision substantially changes the predicted signal: for $\gagg=10^{-12}\,$GeV$^{-1}$ at $E=50\,$keV and the same source geometry, as in Fig.~\ref{fig:Pbar}, our implementation of the
original G20 model gives a low-mass disk-averaged probability of a few
$\times10^{-2}$, about two orders of magnitude above the revised model,
and would strengthen the low-mass coupling limit by a factor of about three.

In local cylindrical coordinates $(r',\phi',z')$ centered on each
structure, we write the field as the curl of a vector potential,
\begin{align} \label{app:modified_B_field}
    \mathbf B
    &= \nabla\times
    \left(A_{z'}\hat{\mathbf z}'\right),
    \\
    A_{z'}
    &= B_1\rho\,W(r',\phi',z')\sin\Phi, \qquad
    \Phi
    = \phi'+\frac{1}{\eta}
    \ln\!\left(\frac{r'+b_0}{\rho+b_0}\right),
\end{align}
where $\rho$ is the reference radius of the structure, $b_0=0.02\rho$ regularizes the logarithmic phase, and $\eta$ sets the pitch of the spiral. The choice $W=e^{-z'^2/H_c^2}$, with $H_c=\rho/\sqrt{\ln2}$, recovers Eq.~(42) of G20. We adopt the parameter values of G20, with $\eta=0.77$ for the CND, and $\eta=0.5$ for the other MCs.

We confine the field by multiplying $W$ by gates based on those of the
gas. Since the gates act on the vector potential, $\nabla\cdot\mathbf B=0$
is preserved, but their gradients introduce additional field contributions in the
transition layers, associated with the confining currents. These scale as
$1/w$ for a transition width $w$. Confinement also changes the mean
strength at fixed $B_1$. We therefore normalize each component to the
reference mean field $B_{\rm ref}$ in Table~4 of G20
($B_{\rm eq}$ for the CND),
\begin{equation} \label{app:B1_normalization}
    B_1=\frac{B_{\rm ref}}{\mathcal A},
    \qquad
    \mathcal A
    =\frac{\int \rd^3x\,G_A|\mathbf B(B_1=1)|}
    {\int \rd^3x\,G_A},
\end{equation}
where $G_A$ is the gas gate of the structure, Eq.~\eqref{app:cloud_gate}
for the MCs and Eq.~\eqref{app:CND_total_gate} for the CND, so that the
average includes the transition regions.

For the MCs, we multiply the G20 profile by a radial
logistic gate with the functional form and width of the gas gate,
Eq.~\eqref{app:cloud_gate}. For the CND, in the inclined frame defined
above, we set $\rho=5\,$pc and
\begin{equation}
    W_{\rm CND}
    =e^{-z'^2/H_c^2}\,
     T(r')\,g_r(r')\,G_{{\rm CND},z}(r',z').
\end{equation}
The taper $T(r')=S(r'/0.5\,{\rm pc})$, with $S$ of
Eq.~\eqref{app:smoothstep}, is the regularization of G20 at $r'<\rho/10$,
which removes the divergence of the $1/r'$ field on the axis.
$G_{{\rm CND},z}$ is the vertical gate of the gas,
Eq.~\eqref{app:CND_vertical_gate}, while the radial gate of
Eq.~\eqref{app:CND_radial_gate} is given a floor $\varepsilon_{\rm in}$
inside the inner edge,
\begin{equation}
    g_r(r')=
    \left[
    \varepsilon_{\rm in}
    +\frac{1-\varepsilon_{\rm in}}
    {1+e^{(r'_{\rm in}-r')/w_r}}
    \right]
    \frac{1}{1+e^{(r'-r'_{\rm out})/w_r}},
\end{equation}
so that part of the CND field, with the G20 spiral shape, extends into
the CC and its minispiral. We adopt $\varepsilon_{\rm in}=0.18$, which gives
a median (mean) plane-of-sky field of $\approx2.4\,(3.0)\,$mG in the modeled
minispiral, near the lower end of the measured interval $2$--$5\,$mG~\cite{Akshaya_2024}, and of $\approx0.1\,(1.1)\,$mG over the whole CC.

Although the fiducial field strengths are compatible with the
observational estimates cited above, the spatial confinement and extension
into the cavity are not uniquely determined by these data. We assess the
resulting uncertainty by varying the confinement parameters.

\begin{figure*}[t]
    \centering
    \includegraphics[width=0.49\textwidth]
        {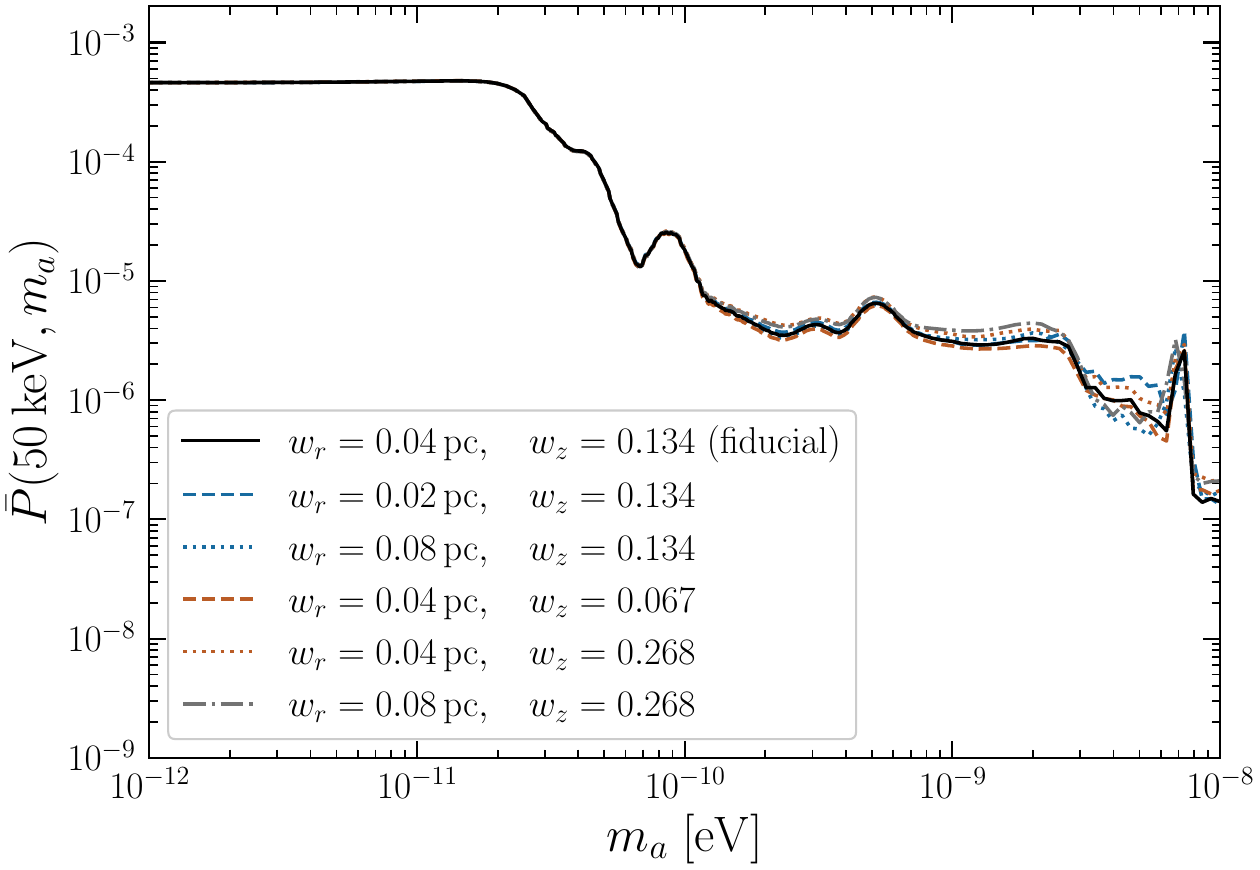}
    \hfill
    \includegraphics[width=0.49\textwidth]
        {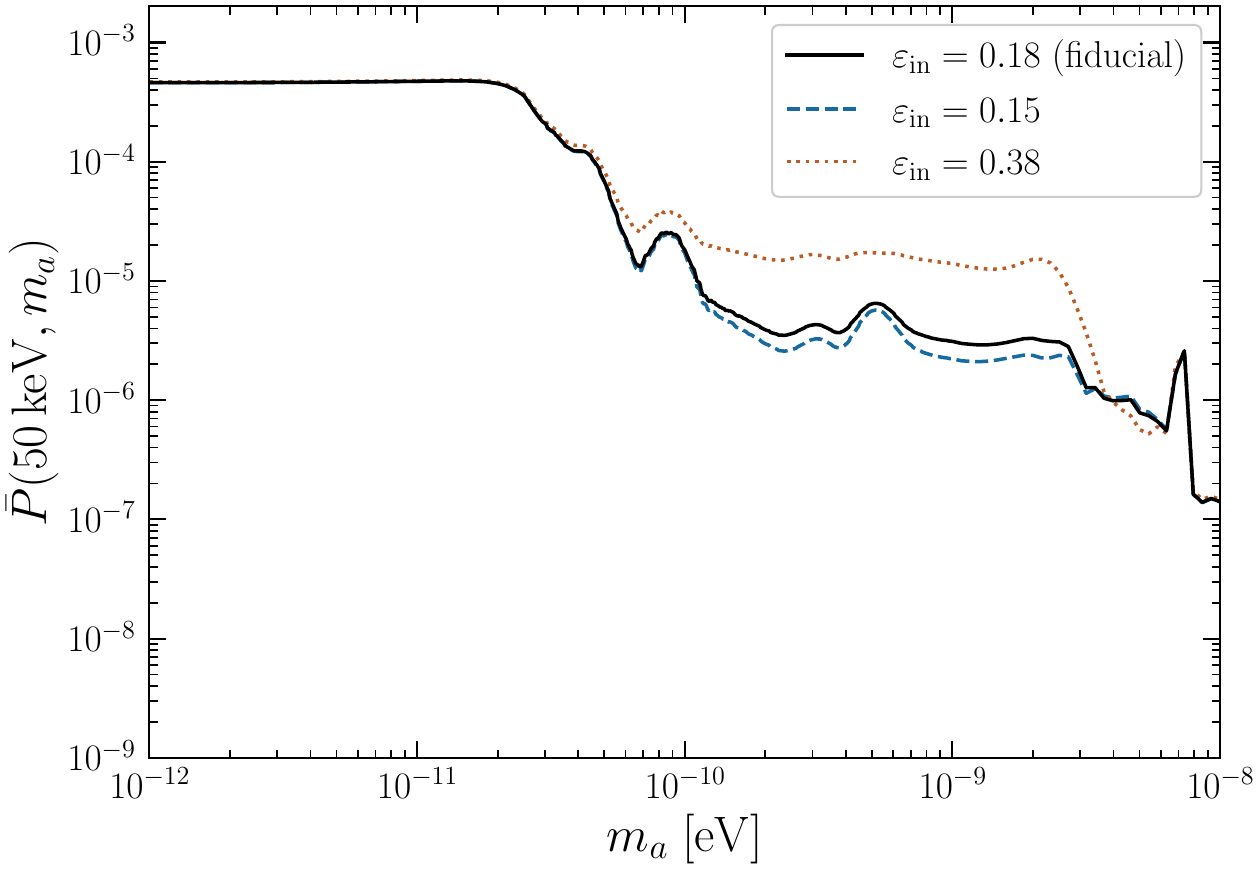}
    \caption{
    Sensitivity of the disk-averaged conversion probability
    to the CND confinement parameters, at $E=50\,$keV and
    $g_{a\gamma\gamma}=10^{-12}\,\mathrm{GeV}^{-1}$,
    averaged within a projected radius of $0.5\,$pc.
    For $m_a\le10^{-9}\,$eV, all curves are averaged over $\pm0.05$ in
    $\log_{10}m_a$, as in Fig.~\ref{fig:Pbar}.
    \emph{Left:} the radial and vertical transition widths
    are halved or doubled relative to their fiducial values,
    keeping $\varepsilon_{\rm in}=0.18$.
    Widths not explicitly varied remain fiducial.
    \emph{Right:} $\varepsilon_{\rm in}=0.15$ and $0.38$, which give the measured median field of $2$--$5\,$mG in the minispiral~\cite{Akshaya_2024}.
    In every case, $B_1$ is recomputed with
    Eq.~\eqref{app:B1_normalization}.
    }
    \label{fig:cnd_confinement_tests}
\end{figure*}

We halve or double the radial and vertical
transition widths and, separately, take $\varepsilon_{\rm in}=0.15$ and $0.38$, the ends of the measured $2$--$5\,$mG interval, recomputing $B_1$ each time (Fig.~\ref{fig:cnd_confinement_tests}).
The probabilities coincide at low masses. At higher masses, the widths change
them by up to a factor of about $2.5$ at individual masses, and
$\varepsilon_{\rm in}$, which sets the field in the CC, by up to a factor of
about four for $10^{-10}\,{\rm eV}\lesssim m_a\lesssim2\times10^{-9}\,$eV.
Since the limit scales as $\bar P^{-1/4}$, the effect on $\gagg$ is much
smaller, as shown in Sec.~\ref{sec:results} and Appendix~\ref{app:sensitivity_tests}.

\subsubsection*{Connection to the large-scale Galactic field}

We connect the central model to the regular JF12 field, including its disk, toroidal-halo and X-shaped components~\cite{JanssonFarrar2012}, through a homogeneous background field $\mathbf B_{\rm bg}$ directed along $\hat{\mathbf z}$. We take $|\mathbf B_{\rm bg}|=10\,\mu$G (Sec.~\ref{sec:magnetic_field}), while the effect of other values is discussed in Sec.~\ref{sec:Pbar}.

\begin{figure}[htb]
    \centering
    \includegraphics[width=0.65\linewidth]{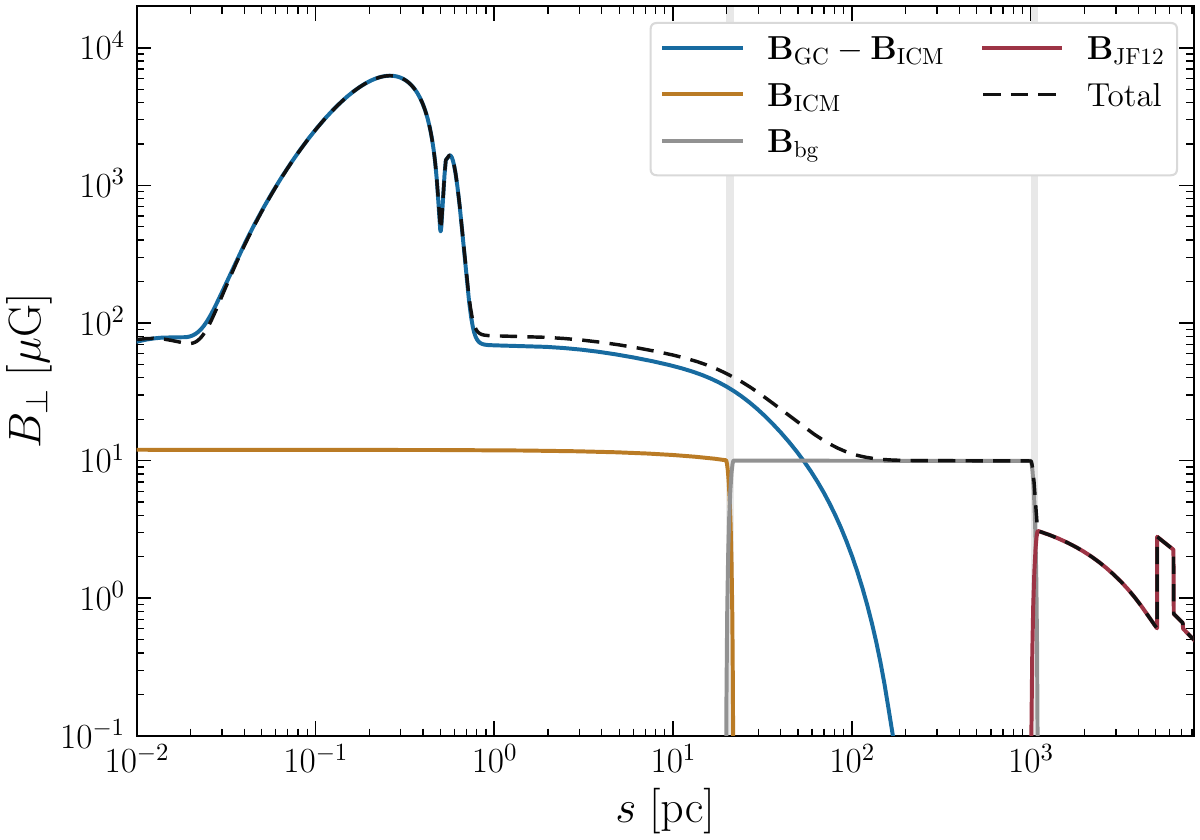}
    \caption{Weighted transverse magnetic-field contributions along the central line of sight from Sgr~A$^\star$ to the observer. The colored curves show $\mathbf B_{\rm GC}-\mathbf B_{\rm ICM}$ (Eq.~\eqref{app:B_GC}), $\mathbf B_{\rm ICM}$, $\mathbf B_{\rm bg}$ and $\mathbf B_{\rm JF12}$, each multiplied by its transition weight; the black curve is the magnitude of their transverse vector sum. The short shaded bands indicate the transitions.}
    \label{fig:field_join}
\end{figure}

The connection uses two transitions, both with the smoothstep of Eq.~\eqref{app:smoothstep} in the spherical Galactocentric distance $r_{\rm sph}=\sqrt{x^2+y^2+z^2}$. Around $r_{\rm sph}\simeq21\,$pc, where the ICM field has decreased to $10\,\mu$G, we replace it with $\mathbf B_{\rm bg}$, while the other components of the central model are kept. For the alternative values of $|\mathbf B_{\rm bg}|$ considered in Sec.~\ref{sec:Pbar}, we change the amplitude of the background field and, for $50\,\mu$G, also the region where it replaces the ICM field. With $3\,\mu$G, it replaces the ICM field from $r_{\rm sph}\simeq21\,$pc to $1\,$kpc, as in the fiducial case. With $50\,\mu$G, it replaces the ICM field everywhere within $r_{\rm sph}\simeq400\,$pc, and decreases to $10\,\mu$G between $380$ and $420\,$pc with the same smoothstep. Between $1$ and $1.1\,$kpc, we replace the entire central field with JF12, which is set to zero inside $1\,$kpc. We note that unlike the confinement of the CND and MC fields, these transitions do not preserve $\nabla\cdot\mathbf B=0$; they only make the field continuous along the lines of sight. Figure~\ref{fig:field_join} shows the resulting contributions to the transverse field along the central line of sight.

\section{NuSTAR data analysis}
\label{app:data_analysis}
This appendix gives the details of the data used in Sec.~\ref{sec:xray_analysis}: the observations and their reduction (Appendix~\ref{app:nustar_data}), and the extraction regions and the background (Appendix~\ref{app:nustar_regions}). Ratios of limits always refer to the fiducial analysis of Sec.~\ref{sec:nustar_analysis}.

\subsection{Observations and data reduction}
\label{app:nustar_data}

In Tab.~\ref{tab:nustar_obs} we list the observations considered. Sgr~A$^\star$ is within $4'$ of the pointing direction in all of them. We reduce the data with the standard NuSTAR pipeline (\texttt{nupipeline}, part of the NuSTARDAS package distributed with HEASoft)~\cite{NuSTARDASguide}, which calibrates the energy and the sky position of each event with the calibration database (CALDB), removes the events on bad pixels, and excludes the time intervals of passages through the South Atlantic Anomaly, Earth occultations and spacecraft slews.
After this first data reduction, we exclude the intervals of the flares documented in Ref.~\cite{Barriere2014}, $6.2\,$ks in ObsID 30001002001 and $1.2\,$ks in 30001002004. Once these flares are removed, we search for undocumented flares of Sgr~A$^\star$ in the $3$--$20\,$keV light curves of a $50''$ region around it. Two observations show anomalies. In ObsID 30001002001, the FPMA count rate decreases by up to about $70\%$ for about $100\,$ks, over the whole detector and at all energies, while the FPMB rate stays constant. To our knowledge, this decrease is not discussed in previous analyses of these data, so we have no criterion to define an interval to exclude. We keep it, after having verified that excluding it changes the limit by $-0.3\%$ to $+2.3\%$ over the whole mass range. In ObsID 30502005004, right after an Earth occultation, the $3$--$20\,$keV rate of the ON region is about $75\%$ higher than average in both modules for about $1.3\,$ks, while that of the OFF annulus is almost unchanged. To our knowledge no such excess, which would be compatible with the end of a Sgr~A$^\star$ flare started during the occultation, is reported in the literature. Its contribution at $30$--$70\,$keV is not detectable, and any residual would be absorbed by the continuum normalization $K_d$ of the dataset, so we keep this interval as well. The light curves of the other observations show no comparable variability. The separate normalizations $K_d$ also absorb the variability between different observations.

\begin{table}[t]
\caption{\label{tab:nustar_obs}NuSTAR observations used in the analysis, in order of decreasing exposure:
start date, exposure of the FPMA and FPMB spectra after the flare selection,
and angular distance between the mean pointing direction and Sgr~A$^\star$
(the off-axis angle of Sgr~A$^\star$ differs by up to about $0.5'$ between
the two modules).}
\begin{ruledtabular}
\begin{tabular}{lcccc}
ObsID & Date & $t_{\rm A}$ [ks] & $t_{\rm B}$ [ks] & Offset [$'$] \\
\hline
30502005004 & 2019-03-29 & 153.0 & 152.8 & 2.3 \\
30001002001 & 2012-07-20 & 149.0 & 147.9 & 2.5 \\
90501330002 & 2019-07-05 & 104.2 & 103.4 & 2.1 \\
30502014002 & 2019-08-13 & 100.2 &  99.9 & 2.1 \\
30302006002 & 2017-04-06 &  65.0 &  65.1 & 1.8 \\
30502014010 & 2021-05-24 &  55.0 &  54.6 & 2.2 \\
30502014006 & 2021-03-30 &  52.6 &  52.1 & 3.3 \\
30302007004 & 2017-07-25 &  50.5 &  50.3 & 2.4 \\
30001002004 & 2012-10-16 &  49.0 &  48.5 & 2.3 \\
40010002001 & 2012-10-13 &  24.5 &  24.2 & 1.0 \\
40010001002 & 2012-10-13 &  24.1 &  23.8 & 3.9 \\
\hline
Total & & 827.1 & 822.6 & \\
\end{tabular}
\end{ruledtabular}
\end{table}

\subsection{Extraction regions and background}
\label{app:nustar_regions}

\begin{figure*}[t]
    \centering
    \includegraphics[width=\textwidth]{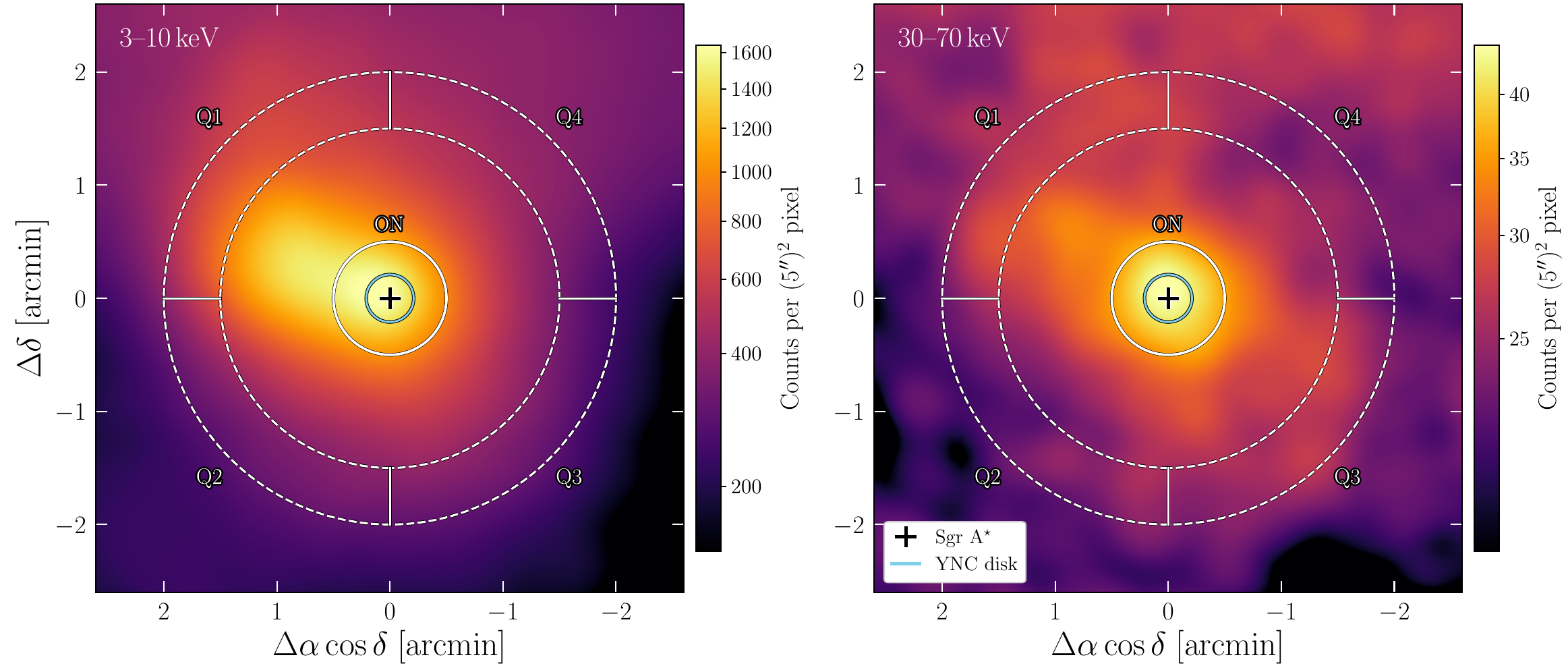}
    \caption{Stacked counts of the eleven NuSTAR observations (FPMA and FPMB) at
    $3$--$10\,$keV (left) and $30$--$70\,$keV (right), smoothed with a Gaussian to better display the emission morphology. White lines: ON region ($30''$) and OFF annulus ($90''$--$120''$),
    divided into the quadrants Q1--Q4. Cyan circle: projected YNC disk
    ($12.6''$). At
    $3$--$10\,$keV the brightest extended emission lies about $1'$ north-east of
    Sgr~A$^\star$, between the ON region and the OFF annulus. At $30$--$70\,$keV
    the emission is concentrated in the ON region, with a
    fainter excess toward the north-east quadrant Q1.}
    \label{fig:sky_regions}
\end{figure*}

The likelihood of Eq.~\eqref{eq:likelihood} treats the 22 datasets separately, but their stacked counts illustrate the spatial distribution of the emission. Figure~\ref{fig:sky_regions} shows them at $3$--$10\,$keV and $30$--$70\,$keV, with the ON region and the OFF annulus. In both bands the emission in the annulus is not uniform and is higher toward the north-east. To quantify this non-uniformity, we divide the annulus into four quadrants and rescale the OFF counts of each quadrant to the ON region, as we do for the full annulus in the fiducial analysis. 
Summed over the 22 datasets, the background that the north-eastern quadrant (Q1) predicts in the ON region at $30$--$70\,$keV is about $10\%$ higher than the mean of the other three quadrants. Q1 is above this mean in 20 of the 22 individual datasets. These were taken at two roll angles that differ by about $180^\circ$, so a structure fixed on the detector would fall in Q1 in one group and in Q3 in the other. The excess is therefore fixed on the sky.

We repeat the analysis using each quadrant alone as OFF region. Over the whole mass range, the limit changes by $-8\%$ to $-19\%$ for Q1, and by $-5\%$ to $+6\%$ for the other three ($-13\%$/$+4\%$ at the three benchmark masses of Fig.~\ref{fig:systematics}). A higher OFF background leaves less room for a signal and makes the limit stronger. The full annulus, which includes Q1, is therefore conservative if the background under the ON region resembles Q1, and too aggressive by at most $6\%$ if it resembles the other three quadrants. Part of this spread is statistical, since each quadrant contains a quarter of the OFF counts. A linear gradient across the field cancels in the concentric annulus, so only structures on the scale of a quadrant matter. We cannot establish which quadrant better represents the background under the ON region, so we keep the annulus chosen a priori and treat the spread of the quadrants as a sensitivity test (Appendix~\ref{app:sensitivity_tests}).

\section{Sensitivity tests}
\label{app:sensitivity_tests}

\begin{figure*}[t]
    \centering
    \includegraphics[width=\textwidth]{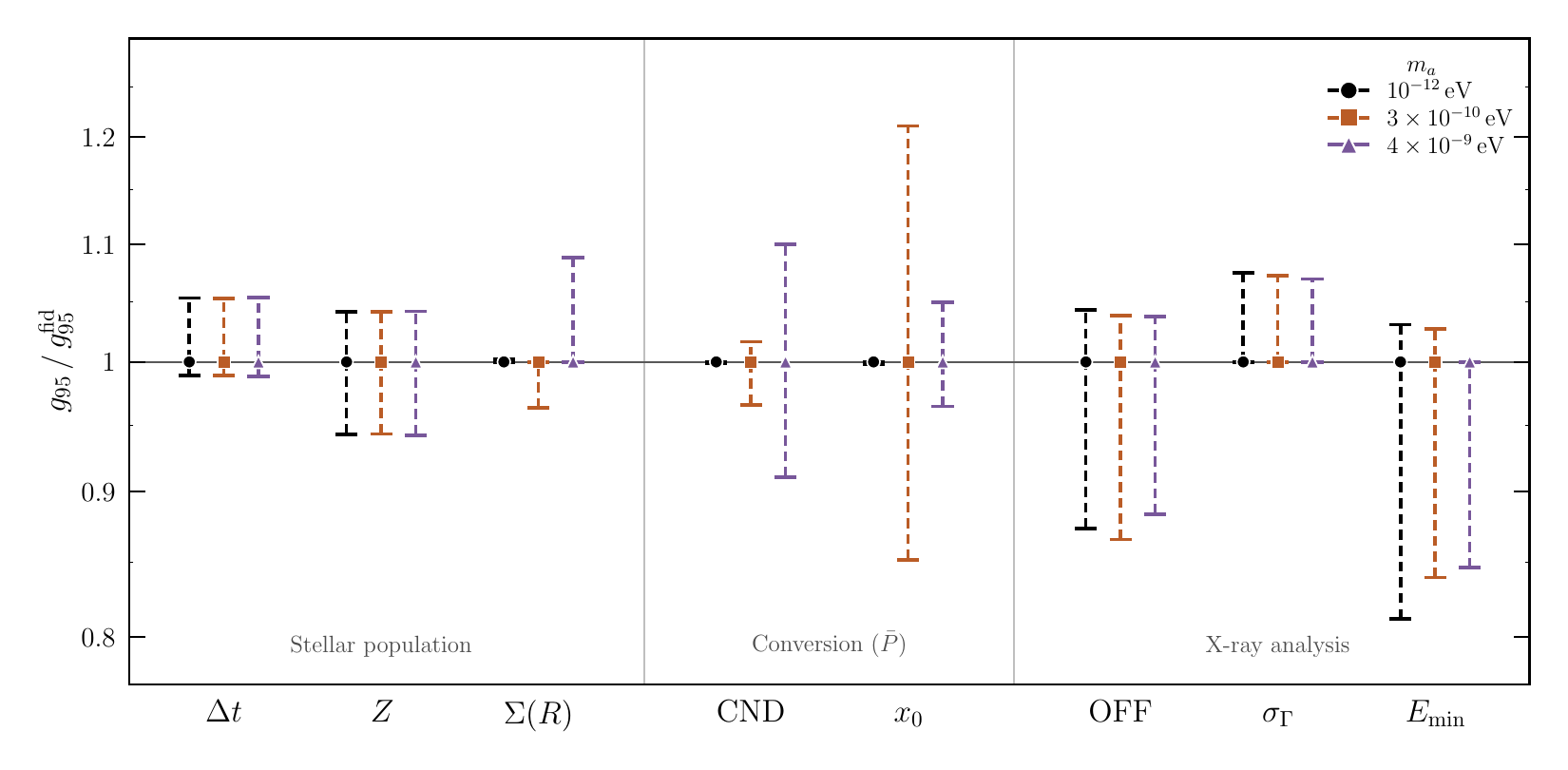}
    \caption{Ratio of the $95\%$ upper limit to the fiducial one, $g_{95}/g_{95}^{\rm fid}$, for the sensitivity tests, in which a fiducial assumption is replaced by alternatives, one at a time, at the three benchmark masses, as in Fig.~\ref{fig:systematics} (colors). Each bar spans the extreme values of the ratio. \emph{Stellar population}: burst duration $\Delta t=0.1$, $0.5$ and $2\,$Myr instead of $1\,$Myr; metallicity $Z=0.01$ and $0.04$ instead of $0.02$; the observed profile $\Sigma(R)$ of the overall population of young stars~\cite{Do2013} instead of a uniform disk. \emph{Conversion probability}: the edges of the CND, with radial and vertical widths halved or doubled; the position $x_0$ of the YNC along the line of sight, $\pm0.25$ and $\pm0.5\,$pc instead of $0$. \emph{X-ray analysis}: each quadrant of the OFF annulus alone; $\sigma_\Gamma=1$ instead of $0.5$; lower edge of the band $E_{\rm min}=25$ or $35\,$keV instead of $30\,$keV.}
    \label{fig:sensitivity_tests}
\end{figure*}

In this appendix we test the signal-model assumptions and analysis choices that, unlike the inputs varied in Sec.~\ref{sec:results}, have no proper observational interval. We replace each of them with one or more alternatives, one at a time, and repeat the NuSTAR data analysis at the three benchmark masses of Fig.~\ref{fig:systematics}. We emphasize that these tests do not define a confidence interval, rather they are meant to illustrate how each ingredient affects the final bounds, and at which masses (Fig.~\ref{fig:sensitivity_tests}).

For the stellar population, the duration of star formation changes the limit by $-1\%$ to $+5\%$ for $\Delta t=0.1$--$2\,$Myr, and the metallicity by $-6\%$/$+4\%$ for $Z=0.01$ and $0.04$ (Appendix~\ref{app:burst_average}). Weighting the lines of sight with the observed profile of the young stars, $\Sigma\propto R^{-0.93}$~\cite{Do2013}, instead of a uniform disk, changes the limit by less than $1\%$ on the plateau, by $-3\%$ at $3\times10^{-10}\,$eV and by $+8\%$ at $4\times10^{-9}\,$eV. 

For the conversion probability, we halve or double the radial and vertical transition widths of the CND (Appendix~\ref{app:galactic_center_models}). These transition regions matter only where the conversion relies on resonances: they change the limit by less than $10\%$ at $4\times10^{-9}\,$eV and by less than $4\%$ at lower masses. A rigid displacement of the YNC along the line of sight, by $x_0=\pm0.25$ and $\pm0.5\,$pc, changes the limit by $-15\%$/$+21\%$ at $3\times10^{-10}\,$eV, where the conversion takes place in the central cavity, and by less than $5\%$ elsewhere.

For the X-ray analysis, using a single quadrant of the OFF annulus changes the limit by $-13\%$/$+4\%$ at the benchmark masses (Appendix~\ref{app:nustar_regions}), relaxing the prior on the photon index to $\sigma_\Gamma=1$ weakens it by $\sim 7\%$, and moving the lower edge of the band to $25\,$keV strengthens it by $15$--$19\%$, while $E_{\rm min}=35\,$keV changes it by at most $3\%$.

\clearpage
\twocolumngrid
\hypersetup{urlcolor=black}
\bibliographystyle{apsrev4-2}
\bibliography{references}

\end{document}